\documentclass[11pt,a4paper]{article}

\usepackage{amsmath,amssymb}     
\usepackage{color}
\usepackage{graphicx}
\usepackage{subfig}
\usepackage{cite}                
\usepackage{hyperref}            
\usepackage{multirow,makecell}
\usepackage{braket,bm}
\usepackage{cancel}
\RequirePackage{doi}
\usepackage{hyperref}
\usepackage{leftidx}
\usepackage[text={17cm,24.5cm},centering]{geometry}
\usepackage{textcomp}
\usepackage{wasysym}
\usepackage{mathtools}

\numberwithin{equation}{section}   

\def \be {\begin{equation}}
\def \ee {\end{equation}}
\def \ba {\begin{array}}
\def \ea {\end{array}}
\def \bea{\begin{eqnarray}}
\def \eea{\end{eqnarray}}

\def \a {\alpha}

\def \G {\Gamma}
\def \d {\delta}
\def \D {\Delta}
\def \dg {\dagger}
\def \e {\epsilon}

\def \m {\mu}

\def \l {\lambda}

\def \s {\sigma}

\def \t {\tau}

\def \mE {\mathcal E}

\def \mN {\mathcal N}
\def \mO {\mathcal O}
\def \mP {\mathcal P}
\def \mQ {\mathcal Q}
\def \mR {\mathcal R}

\def \mT {\mathcal T}

\def \mZ {\mathcal Z}

\def \p {\partial}

\def \lt {\left}
\def \rt {\right}

\def \inf {\infty}

\def \Re {{\textrm{Re}}}

\def \tr {\textrm{tr}}
\def \Tr {{\textrm{Tr}}}

\def \and {{\textrm{and}}}

\begin{document}
\begin{titlepage}
	
	\title{\textbf {Dynamics of charge imbalance resolved negativity after a global quench in free scalar field theory}}
	\author{Hui-Huang Chen\footnote{chenhh@jxnu.edu.cn}~,}
	\date{}
	
	\maketitle
	\underline{}
	\vspace{-12mm}
	
	\begin{center}
		{\it
             College of Physics and Communication Electronics, Jiangxi Normal University,\\ Nanchang 330022, China\\
		}
		\vspace{10mm}
	\end{center}
	\begin{abstract}
	 In this paper, we consider the time evolution of charge imbalance resolved negativity after a global quench in the 1+1 dimensional complex Klein-Gordon theory. We focus on two types of global quenches which are called boundary state quench and mass quench respectively. We first study the boundary state quench where the post-quench dynamic is governed by a massless Hamiltonian. In this case, the temporal evolution of charged imbalance resolved negativity can be obtained first by evaluating the correlators of the fluxed twist field in the upper half plane and then applying Fourier transformation. We test our analytical formulas in the underlying lattice model numerically. We also study the mass quench in the complex harmonic chain where the system evolves according to a massive Hamiltonian after the quench. We argue that our results can be understood in the framework of quasi-particle picture.
	\end{abstract}
	
\end{titlepage}

\thispagestyle{empty}

\newpage

\tableofcontents
\section{Introduction}
In the past two decades, the investigation of various entanglement measures has greatly sharpened our understanding of quantum many-body system, quantum field theory and quantum gravity. In condensed matter theory, entanglement is a powerful tool to characterize different phases of matter \cite{Amico:2007ag, Calabrese:2009qy, Eisert:2008ur}. In the AdS/CFT correspondence, the Ryu-Takayanagi formula \cite{Nishioka:2009un, Ryu:2006bv} firstly opens the route of understanding spacetime from entanglement and this idea turns out to have a key role in the black hole information loss paradox \cite{Hawking:1974sw, Hawking:1976ra, Almheiri:2020cfm}. Entanglement is also an important concept in the studies of equilibration and thermalization of isolated quantum systems. Among all these progress, entanglement entropy is the most successful entanglement measure to characterize the bipartite entanglement of a pure state. When the system is prepared in a pure state $\ket{\psi}$, the reduced density matrix (RDM) of subsystem $A$ is defined by tracing out its complement $B$, $\rho_A=\tr_B\ket{\psi}\bra{\psi}$. From the moments of $\rho_A$, i.e. $\Tr\rho_A^n$, one can obtain the Von Neumann entropy through the replica trick \cite{Calabrese:2004eu}
\be
S\equiv-\Tr(\rho_A\log\rho_A)=\lim_{n\rightarrow 1}S_n
\ee
where $S_n$ is the R\'enyi entropies
\be
S_n=\frac{1}{1-n}\log\Tr\rho_A^n.
\ee
\par Now suppose we are interested in the entanglement between two subsystems $A_1$ and $A_2$, which are not necessarily complementary to each other. In this situation, $\rho_{A_1\cup A_2}$ is general a mixed state, and Von Neumann entropy is no longer a good measure of entanglement. Among different proposals, a computable measure of mixed state entanglement, entanglement negativity (or logarithmic negativity equivalently) turns out to be very useful \cite{Peres:1996dw, vidal2002computable, plenio2005logarithmic}. The definition is
\be
\mathcal{N}=\frac12(||\rho_A^{T_2}||-1),
\ee
where $||O||=\Tr\sqrt{O^{\dg}O}$ denotes the trace norm of the operator $O$ and $\rho_A^{T_2}$ is the partial transpose of RDM $\rho_A$ with respect to degree of freedom of subsystem $A_2$. Let $\ket{e_i^{(1)}}$ and $\ket{e_j^{(2)}}$ be two arbitrary bases of the Hilbert spaces associated to the degree of freedom on $A_1$ and $A_2$ respectively. The partial transpose (with respect to the second space) of $\rho_A$ is defined as
\be
\bra{e_i^{(1)}e_j^{(2)}}\rho_A^{T_2}\ket{e_k^{(1)}e_l^{(2)}}=\bra{e_i^{(1)}e_l^{(2)}}\rho_A\ket{e_k^{(1)}e_j^{(2)}}.
\ee
It's then useful to define the R\'enyi negativity
\be
\mathcal{N}_n=\tr(\rho_A^{T_2})^n
\ee
which could be analytically continued from an even integer $n_e$ to obtain the negativity, using $||\rho_A^{T_2}||=\lim_{n_e\rightarrow 1}\mathcal{N}_{n_e}$. Entanglement negativity has been studied extensively in both quantum field theories (QFT)\cite{calabrese2012entanglement, calabrese2013entanglement, kulaxizi2014conformal, bianchini2016branch, blondeau2016universal, Castro-Alvaredo:2019irt} and in holographic theories \cite{Chaturvedi:2016rcn, Malvimat:2017yaj, chaturvedi2018holographic, kudler2019entanglement}.
\par In recent years, people are interested in the interplay between symmetries and entanglement. When our system exhibit a global symmetry, the entanglement will split into different sectors characterized by eigenvalues of some charge operator. This symmetry resolution of entanglement attract much attention recently \cite{Belin:2013uta, Goldstein:2017bua, PhysRevLett.121.150501, Murciano:2019wdl, Murciano:2020vgh, Horvath:2020vzs, Chen:2021pls, Capizzi:2021zga, Capizzi:2021kys, Zhao:2020qmn, Weisenberger:2021eby, zhao2022charged, Calabrese:2020bys, Capizzi:2022jpx, Ghasemi:2022jxg}. In the context of mixed states, the charge imbalance resolved negativity were also studied in several circumstances \cite{cornfeld2018imbalance, Murciano:2021djk, Chen:2021nma}. However, much fewer results are available for its evolution under non-equilibrium setups \cite{feldman2019dynamics, Parez:2020vsp, Parez:2022xur, Scopa:2022gfw}. It would be very interesting to further investigate the non-equilibrium dynamics of these charge resolved measures of entanglement.
\par In this paper, we will consider the time evolution of the charge imbalance resolved negativity in a special kind of non-equilibrium state known as global quantum quenches. A global quantum quench describes a process in which the sudden change of the Hamiltonian $H_0\rightarrow H$ at a given time that we set as $t=0$, then the initial ground state $\ket{\psi_0}$ of the pre-quench Hamiltonian $H_0$ evolves according to post-quench Hamiltonian $H$. Thus, we have $\ket{\psi(t)}=e^{-iHt}\ket{\psi_0}$. The quench dynamics of various measures of entanglement have been extensively studied in the literature \cite{Calabrese:2005in, Hartman:2013qma, Calabrese:2016xau, Asplund:2015eha, Nozaki:2013vta, Coser:2014gsa, Wen:2015qwa, Cotler:2016acd, MohammadiMozaffar:2018vmk, Zhang:2019kwu, Murciano:2021zvs}. In this paper, we will focus on two types of global quenches in 1+1 dimensional complex Klein-Gordon field theory. The first type is the boundary state quench in which after the quench the mass of the scalar field is zero, so the time evolution is governed by a conformal field theory (CFT). The second type we will consider is a mass quench in the underlying lattice model (complex harmonic chain), where the post-quench dynamic is governed by a massive Hamiltonian.
\par The remaining part of this manuscript is organized as follows. In section \ref{section2}, we briefly review some basic facts about charge imbalance resolution of entanglement negativity. In section \ref{section3}, we discuss how to calculate the dynamics of charged R\'enyi negativity in the boundary state quench setup using CFT techniques. In section \ref{section4}, we will consider the quench dynamics of charged (logarithmic) R\'enyi negativity between two finite intervals in the boundary state quench protocol. In section \ref{section5}, by applying Fourier transformation, we can obtain the charge imbalance resolved negativity from the results derived in the previous section. In section \ref{section6}, we check our analytical predictions against numerical computation in the complex harmonic chain. In section \ref{section7}, we further discuss the mass quench in the underlying lattice model numerically and using quasi-particle interpretation to predict the quench dynamics of charged logarithmic negativity, finding perfect agreement. Finally, we conclude in section \ref{section8} and discuss some possible interesting extensions of this work.
\section{Charge imbalance resolution of negativity}\label{section2}
In this section, we will briefly review the decomposition of entanglement negativity under a global internal symmetry.
\par Let us first review some basic facts about the symmetry resolution of the entanglement entropy. We assume that our system exhibit a global $U(1)$ symmetry generated by a local charge $Q$. If $[\rho, Q]=0$ (this can be achieved if $\rho$ only acts non-trivially on the eigenspace of $Q$), then we have $[\rho_A,Q_A]=0$, which implies that $\rho_A$ admits charge decomposition according to eigenvalues $q$ of local charge $Q_A$
\be
\rho_A=\oplus_q\mathcal{P}_q\rho_A=\oplus_qp(q)\rho_A(q),\quad p(q)=\Tr(\mathcal{P}_q\rho_A),
\ee
where $\mathcal{P}_q$ is the projection operator which projects the space to the eigenspace corresponding to eigenvalue $q$.
The symmetry resolved R\'enyi entropies are defined as
\be
S_n(q)=\frac{1}{1-n}\log\Tr[\rho_A(q)]^n.
\ee
\par It's convenient to first introduce the charged moments of $\rho_A$,
\be\label{Znmiu}
Z_n(\m)=\Tr(e^{i\m Q_A}\rho_A^n).
\ee
Then it's sufficient to compute its Fourier transform
\be
Z_n(q)=\int_0^{2\pi}\frac{d\m}{2\pi}e^{-iq\m}Z_n(\m)
\ee
to obtain the R\'enyi entropies of the sector with charge $q$ as
\be
S_n(q)=\frac{1}{1-n}\log\left[\frac{Z_n(q)}{Z_1(q)^n}\right].
\ee
Finally, the symmetry resolved entanglement entropy can be obtained by taking the replica limit $S(q)=\lim_{n\rightarrow 1}S_n(q)$.
\par Now we briefly review the symmetry decomposition of entanglement negativity under the $U(1)$ charge $Q$. Since the charge $Q$ is local, we can write $Q_A=Q_1+Q_2$, where $Q_1$ and $Q_2$ are the charges corresponding to sub subsystem $A_1$ and $A_2$ respectively. From the relation $[\rho_A,Q_A]=0$, performing a partial transposition with respect to the second region $A_2$ of subsystem $A$, we obtain
\be
[\rho_A^{T_2},\mQ_A]=0,\quad \mQ_A\equiv Q_1-Q_2^{T_2},
\ee
where we have introduced the charge imbalance operator $\mQ_A$ and we will denote its eigenvalues as $\mathrm{q}$ to make a distinction with the eigenvalues of $Q_A$.
Then $\rho_A^{T_2}$ has a block matrix form, each block was characterized by different eigenvalues $\mathrm{q}$ of the imbalance operator $\mQ_A$. If we write
\be
\rho_A^{T_2}=\oplus_{\mathrm{q}}\mathcal{P}_{\mathrm{q}}\rho_A^{T_2},\quad \rho_A^{T_2}({\mathrm{q}})=\frac{\mathcal{P}_{\mathrm{q}}\rho_A^{T_2}}{\Tr(\mathcal{P}_{\mathrm{q}}\rho_A^{T_2})}.
\ee
Then we have
\be
\rho_A^{T_2}=\oplus_{\mathrm{q}} p({\mathrm{q}})\rho_A^{T_2}({\mathrm{q}}),
\ee
where $p({\mathrm{q}})=\Tr(\mP_{\mathrm{q}}\rho_A^{T_2})$ is the probability of finding ${\mathrm{q}}$ as the outcome of measurement of $\mQ_A$.
\par The charge imbalance resolved negativity is defined as
\be
\mathcal{N}({\mathrm{q}})=\frac{1}{2}(\Tr|\rho_A^{T_2}({\mathrm{q}})|-1).
\ee
The total negativity is given by the sum of charge imbalance resolved negativity weighted by the corresponding probability
\be\label{Recon}
\mathcal{N}=\sum_{\mathrm{q}}p({\mathrm{q}})\mathcal{N}({\mathrm{q}}).
\ee
It's useful to define the charge imbalance resolved R\'enyi negativity
\be
\mN_n(\mathrm{q})=\frac{N_n(\mathrm{q})-1}{2},\quad N_n({\mathrm{q}})=\Tr[(\rho_A^{T_2}({\mathrm{q}}))^n]=\frac{1}{p({\mathrm{q}})^n}\Tr[\mathcal{P}_{\mathrm{q}}(\rho_A^{T_2})^n].
\ee
Then the charge imbalance entanglement negativity can be obtained by taking the limit
\be
\mathcal{N}({\mathrm{q}})=\lim_{n_e\rightarrow 1}\mN_{n_e}(\mathrm{q}).
\ee
The projection operator $\mathcal{P}_\mathrm{q}$ has the following integral representation
\be
\mathcal{P}_{\mathrm{q}}=\int_{0}^{2\pi}\frac{d\m}{2\pi}e^{-i\m {\mathrm{q}}}e^{i\m\mathcal{Q}_A}.
\ee
It's convenient to first introduce the charged R\'enyi negativity
\be\label{Rnmu}
R_n(\m)=\Tr[(\rho_A^{T_2})^ne^{i\m\mathcal{Q}_A}],
\ee
and it's Fourier transformation
\be
\mathcal{Z}_{T_2,n}(\mathrm{q})=\int_0^{2\pi}\frac{d\m}{2\pi} e^{-i\m\mathrm{q}}R_n(\m)
\ee
then the charge imbalance resolved R\'enyi negativity are related by Fourier transform
\be\label{pq}
\mN_n({\mathrm{q}})=\frac{\mathcal{Z}_{T_2,n}(\mathrm{q})}{p(\mathrm{q})^n},\qquad p(\mathrm{q})=\int_0^{2\pi}\frac{d\m}{2\pi} e^{-i\m {\mathrm{q}}}R_1(\m).
\ee
It's also useful to introduce the charged R\'enyi logarithmic negativity $\mathcal{E}_n(\m)$, defined as
\be
\mathcal{E}_n(\m)=\log\Tr[(\rho_A^{T_2})^ne^{i\m\mathcal{Q}_A}].
\ee
The charged logarithmic negativity is defined by taking the following replica limit
\be
\mathcal{E}(\m)=\lim_{n_e\rightarrow 1}\mathcal{E}_{n_e}(\m).
\ee
Then the charge imbalance resolved negativity can be obtained as
\be\label{ZT2q}
\mathcal{N}(\mathrm{q})=\frac12\left(\frac{\mathcal{Z}_{T_2}(\mathrm{q})}{p(\mathrm{q})}-1\right),\quad \mathcal{Z}_{T_2}(\mathrm{q})\equiv\lim_{n_e\rightarrow 1}\mathcal{Z}_{T_2,n_e}(\mathrm{q})
=\int_0^{2\pi}\frac{d\m}{2\pi} e^{-i\m\mathrm{q}}e^{\mE(\m)}.
\ee
In section \ref{section5}, we will use eq.~(\ref{ZT2q}) and eq.~(\ref{pq}) to evaluate the charge imbalance resolved negativity.
\section{Boundary state quench}\label{section3}
In this paper, we will consider the 1+1 dimensional complex free scalar field theory with the Euclidean action given by
\be
\mathcal{A}=\int d^2x(\partial_{\mu}\phi^{\dg}\partial_{\mu}\phi+m^2\phi^{\dg}\phi).
\ee
This action exhibit a $U(1)$ symmetry, i.e. the phase transformation of the field $\phi\rightarrow e^{i\theta}\phi, \phi^{\dg}\rightarrow e^{-i\theta}\phi^{\dg}$ leaves the action invariant. The Hamiltonian of this theory is
\be
H=\int dx(\pi^{\dg}\pi+\partial_x\phi^{\dg}\partial_x\phi+m^2\phi^{\dg}\phi)
\ee
with $\pi$ is the canonical momentum for $\phi$.
\par We can rewrite this theory in terms of two real scalar fields $\phi^{(1)}$ and $\phi^{(2)}$ with $\phi=\frac{1}{\sqrt{2}}(\phi^{(1)}+i\phi^{(2)})$.
The Hamiltonian in terms of these variables becomes
\be
H=\frac12\sum_{\a=1,2}\int d^2x[(\pi^{(\a)})^2+(\partial_x\phi^{(\a)})^2+m^2(\phi^{(\a)})^2].
\ee
The Hamiltonian can be diagonalized in terms of particle and anti-particle mode operators $a(p),a^{\dg}(p)$ and $b(p),b^{\dg}(p)$ with the commutation relation $[a(p),a^{\dg}(q)]=[b(p),b^{\dg}(q)]=2\pi\delta(p-q)$ and all other commutators vanishing. We have
\be
H=\int\frac{dp}{2\pi}e(p)(a^{\dg}(p)a(p)+b^{\dg}(p)b(p))
\ee
with $e(p)=\sqrt{m^2+p^2}$, while the conserved charge corresponding to the global  $U(1)$ symmetry is
\be
Q=\int\frac{dp}{2\pi}(a^{\dg}(p)a(p)-b^{\dg}(p)b(p))=\int dx(a^{\dg}(x)a(x)-b^{\dg}(x)b(x)).
\ee
In the second equal sign of above equation, the conserved charge are expressed as integral of local density in real space. Thus its value in a given subsystem $A$ is the same integral restricted to $A$,
\be
Q_A=\int_A dx(a^{\dg}(x)a(x)-b^{\dg}(x)b(x)).
\ee
\par The lattice version of the complex Klein-Gordon field theory is the complex harmonic chain which is equivalent to two decoupled real harmonic chains.
The Hamiltonian of the real harmonic chain made by $L$ sites reads
\be
H_{HC}=\frac12\sum_{j=0}^{L-1}\lt[\pi_j^2+m^2\phi_j^2+(\phi_{j+1}-\phi_j)^2\rt],
\ee
where periodic boundary conditions $\phi_L\equiv \phi_0,\pi_L\equiv \pi_0$ are imposed and variables $\pi_j$ and $\phi_j$ satisfy standard bosonic commutation relations $[\phi_i,\phi_j]=[\pi_i,\pi_j]=0$ and $[\phi_i,\pi_j]=\mathrm{i}\d_{ij}$.  The lattice version of the complex scalar field theory is the sum of two of the above harmonic chain. In terms of the variables $\phi^{(1)},\pi^{(1)}$ and $\phi^{(2)},\pi^{(2)}$, the Hamiltonian is
\be\label{HCHC}
H_{CHC}=H_{HC}(\pi^{(1)},\phi^{(1)})+H_{HC}(\pi^{(2)},\phi^{(2)}).
\ee
\par The Hamiltonian eq.~(\ref{HCHC}) can be diagonalized by introducing the creation and annihilation operators $a_k,a_k^{\dg}$ and $b_k,b_k^{\dg}$, satisfying $[a_k,a_{k'}^{\dg}]=\d_{kk'}$ and $[b_k,b_{k'}^{\dg}]=\d_{kk'}$. In terms of these operators, the Hamiltonian eq.~(\ref{HCHC}) is diagonal
\be\label{HCHC1}
H_{CHC}=\sum_{k=0}^{L-1}e_k(a_k^{\dg}a_k+b_k^{\dg}b_k),\quad e_k=\sqrt{m^2+4\sin^2\lt(\frac{\pi k}{L}\rt)}.
\ee
While the $U(1)$ charge is
\be\label{Q}
Q=\sum_{k=0}^{L-1}(a_k^{\dg}a_k-b_k^{\dg}b_k).
\ee
The conserved charge is local and can also be written in the position space and for a given subsystem $A$ reads
\be
Q_A=\sum_{j\in A}(a_j^{\dg}a_j-b_j^{\dg}b_j).
\ee
\subsection{The path integral approach to boundary state quench}
In this section, we will focus on the case in which after the quench the mass of the scalar field is zero, so time evolution is governed by a CFT. Let's first briefly review the imaginary time formalism for the global quenches where the post-quench dynamic is govern by a CFT. This quench setup is called a boundary state quench as the correlation measures after quench can be described by boundary conformal field theory (BCFT), where the gapped initial state are represent by a spacetime boundary.
\par The expectation value of a product of equal-time local operators in time dependent state $\ket{\psi(t)}$ is
\be
\langle\mO(t,\{x_i\})\rangle=Z^{-1}\bra{\psi_0}e^{iHt-\tau_0H}\mO(\{x_i\})e^{-iHt-\tau_0H}\ket{\psi_0}.
\ee
where two factors $e^{-\tau_0H}$ have been introduced to make the path integral representation of this expectation value absolutely convergent and $Z=\bra{\psi_0}e^{-2\tau_0H}\ket{\psi_0}$ is the normalization factor. The above prescription is equivalent to assuming that the initial state has the form $\ket{\psi_0}\propto e^{-\tau_0H}\ket{B}$, where $\ket{B}$ is some conformal boundary state. One could interpret $\tau_0$ as being proportional to the correlation length of the initial state. Thus, the predictions made by this approach are expected to be valid only in the spacetime scaling limit, $t\gg\tau_0,|x_i-x_j|\gg\tau_0, t/|x_i-x_j|=\text{finite}$. The above correlator admits a path integral represent
\be
\langle\mO(t,\{x_i\})\rangle=\frac{1}{Z}\int[d\phi(x,\tau)]\mO(\{x_i\},\tau=\tau_0+it)e^{-\int_{\tau_1}^{\tau_2}d\tau L}\braket{\psi_0|\phi(x,\tau_2)}\braket{\phi(x,\tau_1)|\psi_0}.
\ee
where $L$ is the Euclidean Lagrangian corresponding to the post-quench Hamiltonian $H$. We need to identify $\tau_1=0$ and $\tau_2=2\tau_0$. We should compute the path integral considering $\tau$ real and only at the end of the computation to analytically continue it to $\tau=\tau_0+it$.
\subsection{Evolution of charged moments of RDM}
In this section, we briefly review the strategy of computing the charged R\'enyi negativity in two-dimensional CFT (see \cite{Chen:2021nma} for more details).
In a generic two-dimensional QFT, we can view the charged moments $Z_n(\m)$ (defined in eq.~(\ref{Znmiu})) as the partition function on the Riemann surface $\mR_{n,N}$ pierced by an Aharonov-Bohm flux, such that the total phase accumulated by the field upon going through the entire surface is $\m$. The presence of the flux corresponds to impose an additional twist on the boundary of subsystem $A$. This additional twist fuses with the replica (ordinary) twist field at the endpoints of subsystem $A$ and can be implemented by two local fields $\mT_{n,\m}$ and $\tilde{\mT}_{n,\m}$ named as fluxed twist fields and fluxed anti-twist fields. These fluxed twist fields take into account not only the internal permutational symmetry among the replicas but also the presence of the flux. The partition function on the fluxed Riemann surface is thus proportional to the $2N$-point function of these fluxed twist operators.
\par We consider the subsystem $A$ consists of two disjoint intervals on the real axis, $A=A_1\cup A_2$ with $A_1=[u_1,u_2],A_2=[u_3,u_4]$. The charged moments of RDM of vacuum state $\rho_A$, i.e. $\tr\rho_A^ne^{i\m Q_A}$ is equivalent to a four-point function of the fluxed twist fields
\be
\Tr\rho_A^ne^{i\m Q_A}=\langle\mathcal{T}_{n,\m}(w_1)\tilde{\mathcal{T}}_{n,\m}(w_{2})\mathcal{T}_{n,\m}(w_3)\tilde{\mathcal{T}}_{n,\m}(w_{4})\rangle_{\text{strip}}~,\qquad w_i=u_i+i\tau,
\ee
where $w=u+i\tau$ with $\tau\in\mathbb{R},\tau\in(0,2\pi)$ is the coordinate of the strip.
The fluxed twist fields $\mathcal{T}_{n,\m}$ and fluxed anti-twist field $\tilde{\mathcal{T}}_{n,\m}$ are primary operators and they have the same dimension \cite{Belin:2013uta, Murciano:2020vgh}
\be
\D_{n,\m}=\frac{1}{6}\lt(n-\frac{1}{n}\rt)-\frac{\m^2}{4\pi^2n}+\frac{\m}{2\pi n}.
\ee
The charged moments of the partially transposed RDM can be obtained from the correlator above with the fluxed twist field $\mathcal{T}_{n,\m}$ and $\tilde{\mathcal{T}}_{n,\m}$ at the endpoints of $A_2$ exchanged while the remaining ones keep the same, giving
\be
\Tr[(\rho_A^{T_2})^ne^{i\m\mQ_A}]
=\langle\mathcal{T}_{n,\m}(w_1)\tilde{\mathcal{T}}_{n,\m}(w_{2})\tilde{\mathcal{T}}_{n,\m}(w_3)\mathcal{T}_{n,\m}(w_{4})\rangle_{\text{strip}}~.
\ee
By conformal map $z(w)=e^{\pi w/2\tau_0}$, the four point function on the strip are mapped to the upper half plane (UHP), then we can write
\be\label{Ninterval}
\Tr\rho_A^ne^{i\m Q_A}=\left(\frac{\pi}{2\tau_0}\right)^{4\Delta_{n,\m}}|z_1z_2z_3z_4|^{\Delta_{n,\m}}\langle\mathcal{T}_{n,\m}(z_{1})\tilde{\mathcal{T}}_{n,\m}(z_{2})
\mathcal{T}_{n,\m}(z_{3})\tilde{\mathcal{T}}_{n,\m}(z_{4})\rangle_{\text{UHP}},
\ee
By global conformal symmetry, four-point functions of primary fields on the upper half plane depend on 6 cross ratios $\eta_{i,j}$ and have the following structure
\be
\langle\mathcal{T}_{n,\m}(z_{1})\tilde{\mathcal{T}}_{n,\m}(z_{2})\mathcal{T}_{n,\m}(z_{3})\tilde{\mathcal{T}}_{n,\m}(z_{4})\rangle_{\text{UHP}}=c_{n,\m}^2\prod_{a=1}^{4}|z_a-\bar z_a|^{-\D_{n,\m}}\left(\frac{\eta_{1,3}\eta_{2,4}}{\eta_{1,2}\eta_{1,4}\eta_{2,3}\eta_{3,4}}\right)^{\D_{n,\m}}\mathcal{F}_{n}(\{\eta_{j,k}\})
\ee
where the cross ratios are defined by $\eta_{i,j}=\frac{(z_i-z_j)(\bar z_i-\bar z_j)}{(z_i-\bar z_j)(\bar z_i-z_j)}$ and $\mathcal{F}(\{\eta_{j,k}\})$ depend on the full operator content of the theory and usually is very complicated. The constants $c_{n,\m}$ are also non-universal and are known for some specific theories.
\par While for the charged R\'enyi negativity, we have
\be
\begin{split}
\Tr[(\rho_A^{T_2})^ne^{i\m\mQ_A}]=\left(\frac{\pi}{2\tau_0}\right)^{4\Delta_{n,\m}}\prod_{a=1}^4|z_a|^{\Delta_{n,\m}}
\langle\mathcal{T}_{n,\m}(z_{1})\tilde{\mathcal{T}}_{n,\m}(z_{2})\tilde{\mathcal{T}}_{n,\m}(z_{3})\mathcal{T}_{n,\m}(z_{4})\rangle_{\text{UHP}}\\
=c_{n,\m}^2\prod_{a=1}^{4}\Big|\frac{z_a}{z_a-\bar z_a}\Big|^{\D_{n,\m}}\frac{1}{\eta_{1,2}^{\D_{n,\m}}\eta_{3,4}^{\D_{n,\m}}}
\left(\frac{\eta_{1,4}\eta_{2,3}}{\eta_{1,3}\eta_{2,4}}\right)^{\D^{(2)}_{n,\m}/2-\D_{n,\m}}\mathcal{G}_{n}(\{\eta_{j,k}\})
\end{split}
\ee
\par Fortunately, in the spacetime scaling regime, we have $\eta_{i,j}\rightarrow 0,1$ or $\infty$, and the non-universal functions $\mathcal{F}_n(\{\eta_{j,k}\}),\mathcal{G}_n(\{\eta_{j,k}\})$ are just constant. In the following section we will just omit these non-universal functions and just focus on the universal part.
\par There is one important comment we must address. To compute the four-point function of fluxed twist fields on the strip, we apply a conformal mapping from the strip to the UHP. This is somewhat similar to the case of computation of the negativity at finite temperature, where it is well-known that if the partial transposition involves an infinite part of an infinite system at finite temperature, using the conformal map from the cylinder to the complex plane is naively wrong \cite{Calabrese:2014yza}. This is not the case if one, for example, is interested in the negativity between two (adjacent or disjoint) finite intervals \cite{2014Entanglement, Hoogeveen:2014bqa}. In our situation, we are concentrated on the charged R\'enyi negativity between two finite intervals. Thus it's indeed free of these troubles.
\section{Evolution of charged logarithmic negativity in boundary state quench}\label{section4}
In this section, we will consider the dynamics of the charged logarithmic negativity between two intervals after a global quench whose post-quench evolution is governed by a massless Hamiltonian.
\subsection{Bipartite system}
It's convenient to first study the case in which $A=A_1\cup A_2$ is the entire system. In this case, $\rho_A$ corresponds to a pure state. As explained in previous section, in this case, the temporal evolution of the charged R\'enyi negativity is governed by $\langle\mathcal{T}^2_{n,\m}(w_{1})\tilde{\mathcal{T}}^2_{n,\m}(w_{2})\rangle$ on the strip. The strip two-point function can be computed from the one in the UHP which has the standard form
\be
\langle\mathcal{T}^2_{n,\m}(w_{1})\tilde{\mathcal{T}}^2_{n,\m}(w_{2})\rangle_{\text{UHP}}=|(z_1-\bar z_1)(z_2-\bar z_2)\eta_{1,2}|^{-\D_{n,\m}^{(2)}},
\ee
where
\be
\D_{n,\m}^{(2)}=
\begin{cases}
\D_{n,2\m},\quad \text{odd}~n\\
2\D_{\frac{n}{2},\m},\quad \text{even}~n
\end{cases}
\ee
After mapping the correlator in the UHP to the one in the strip, we find
\be
\begin{split}
\Tr[(\rho_A^{T_2})^ne^{i\m\mQ_A}]=\langle\mathcal{T}^2_{n,\m}(w_{1})\tilde{\mathcal{T}}^2_{n,\m}(w_{2})\rangle_{\text{strip}}\\
=\left(\frac{\pi}{2\tau_0}\right)^{2\D_{n,\m}^{(2)}}\prod_{a=1}^{2}\Big|\frac{z_a}{z_a-\bar z_a}\frac{1}{\eta_{1,2}}\Big|^{\D_{n,\m}^{(2)}}.
\end{split}
\ee
In the spacetime scaling limit regime $t\gg\tau_0,|u_i-u_j|\gg\tau_0$, we have
\be
\log\Big|\frac{z_a}{z_a-\bar z_a}\Big|\rightarrow -\frac{\pi t}{2\tau_0},\qquad\log\eta_{i,j}\rightarrow\frac{\pi}{2\tau_0}(|u_i-u_j|-\max(2t,|u_i-u_j|)).
\ee
Then it's straightforward to derive the charged R\'enyi logarithmic negativity
\be
\mathcal{E}_n(\m)=-\frac{\pi\D_{n,\m}^{(2)}}{2\tau_0}\min(2t,l_1).
\ee
In particular, for $n=1$, the result is
\be
\mathcal{E}_1(\m)=-\frac{\pi}{2\tau_0}h_1(2\m)\min(2t,l_1),\qquad h_1(\m)=-\frac{\m^2}{4\pi^2}+\frac{\m}{2\pi}.
\ee
The charged logarithmic negativity is obtained by taking replica limit $n_e\rightarrow 1$ in $\mE_{n_e}(\m)$
\be
\mathcal{E}(\m)=-\frac{\pi}{2\tau_0}h(\m)\min(2t,l_1),\qquad h(\m)=-\frac12-\frac{\m^2}{\pi^2}+\frac{2\m}{\pi}.
\ee

\subsection{Two adjacent intervals}
In this case, the quench dynamics of the charged R\'enyi negativity between two adjacent intervals can be obtained by studying the three point function $\langle\mathcal{T}_{n,\m}(w_{1})\tilde{\mathcal{T}}^2_{n,\m}(w_{2})\mathcal{T}_{n,\m}(w_{2})\rangle$ on the strip which can be computed by conformal mapping from the three point function on the UHP. After dropping the non-universal functions, we finally have
\be
\begin{split}
\Tr[(\rho_A^{T_2})^ne^{i\m\mQ_A}]=\langle\mathcal{T}_{n,\m}(w_{1})\tilde{\mathcal{T}}^2_{n,\m}(w_{2})\mathcal{T}_{n,\m}(w_{3})\rangle_{\text{strip}}\\
=\left(\frac{\pi}{2\tau_0}\right)^{\D}\prod_{a=1}^{3}\Big|\frac{z_a}{z_a-\bar z_a}\Big|^{\D_{(a)}}
\left(\frac{\eta_{1,3}^{\D_{n,\m}^{(2)}-2\D_{n,\m}}}{\eta_{1,2}^{\D_n^{(2)}}\eta_{2,3}^{\D_n^{(2)}}}\right)^{1/2}.
\end{split}
\ee
where $\D=2\D_{n,\m}+\D^{(2)}_{n,\m}$, $\D_{(1)}=\D_{(3)}=\D_{n,\m}$ and $\D_{(2)}=\D^{(2)}_{n,\m}$. Then the CFT prediction for the time evolution of the charged logarithmic R\'enyi negativity is given by
\be\label{Enmiua}
\mathcal{E}_n(\m)=-\frac{\pi}{4\tau_0}\left[\D_{n,\m}^{(2)}(\min(2t,l_1)+\min(2t,l_2))-(\D_{n,\m}^{(2)}-2\D_{n,\m})\min(2t,l_1+l_2)\right].
\ee
where we have defined $l_1=u_2-u_1,l_2=u_3-u_2$.
\par The temporal evolution of the charged logarithmic negativity is obtained by taking replica limit $n_e\rightarrow 1$ in $\mE_{n_e}(\m)$. The result is
\be\label{Emiua}
\mathcal{E}(\m)=-\frac{\pi}{4\tau_0}\left[h(\m)(\min(2t,l_1)+\min(2t,l_2))-h_2(\m)\min(2t,l_1+l_2)\right].
\ee
with
\be
h_2(\m)=-\frac12-\frac{\m^2}{2\pi^2}+\frac{\m}{\pi}.
\ee
From eq.~(\ref{Enmiua}), we have
\be
\mathcal{E}_1(\m)=-\frac{\pi}{4\tau_0}\left[h_1(2\m)(\min(2t,l_1)+\min(2t,l_2))-h_3(\m)\min(2t,l_1+l_2)\right],
\ee
where $h_3(\m)=-\frac{\m^2}{2\pi^2}$.
\subsection{Two disjoint intervals}
The universal part the charged R\'enyi negativity is
\be
\begin{split}
&\Tr\rho_A^ne^{i\m\mQ_A}=\langle\mathcal{T}_{n,\m}(w_1)\tilde{\mathcal{T}}_{n,\m}(w_{2})\tilde{\mathcal{T}}_{n,\m}(w_3)\mathcal{T}_{n,\m}(w_{4})\rangle_{\text{strip}}\\
&=\left(\frac{\pi}{2\tau_0}\right)^{4\Delta_{n,\m}}\prod_{a=1}^{4}\Big|\frac{z_a}{z_a-\bar z_a}\Big|^{\D_{n,\m}}\frac{1}{\eta_{1,2}^{\D_{n,\m}}\eta_{3,4}^{\D_{n,\m}}}
\left(\frac{\eta_{1,4}\eta_{2,3}}{\eta_{1,3}\eta_{2,4}}\right)^{\D^{(2)}_{n,\m}/2-\D_{n,\m}}.
\end{split}
\ee
In this case, the CFT prediction for the time evolution of the charged logarithmic R\'enyi negativity is given by
\be\label{Enmiud}
\begin{split}
&\mathcal{E}_n(\m)=-\frac{\pi}{2\tau_0}\big[\D_{n,\m}\left(\min(2t,l_1)+\min(2t,l_2)\right)+(\D_{n,\m}^{(2)}/2-\D_{n,\m})\\
&\times(\max(2t,l_1+l_2+d)+\max(2t,d)-\max(2t,l_1+d)-\max(2t,l_2+d))\big].
\end{split}
\ee
where $l_1=u_2-u_1,l_2=u_4-u_3$ and $d=u_3-u_2$ are the length of the subsubsystem $A_1,A_2$ and the distance between them respectively.
\par The charged logarithmic negativity is easily obtained as
\be\label{Emiud}
\begin{split}
&\mathcal{E}(\m)=-\frac{\pi}{2\tau_0}\big[h_1(\m)\left(\min(2t,l_1)+\min(2t,l_2)\right)+\frac12h_2(\m)\\
&\times(\max(2t,l_1+l_2+d)+\max(2t,d)-\max(2t,l_1+d)-\max(2t,l_2+d))\big].
\end{split}
\ee
As before, we also report the expression of $\mathcal{E}_1(\m)$ here since it will be useful in the computation of charge imbalance resolved negativity. The result is
\be
\begin{split}
&\mathcal{E}_1(\m)=-\frac{\pi}{2\tau_0}[h_1(\m)(\min(2t,l_1)+\min(2t,l_2))\\
&+\frac12 h_3(\m)(\max(2t,l_1+l_2+d)+\max(2t,d)-\max(2t,l_1+d)-\max(t,l_2+d))].
\end{split}
\ee
\par We stress that eq.~(\ref{Enmiua}) and eq.~(\ref{Enmiud}) are only valid for CFT in which there is a perfect linear dispersion. However this is not the case for the underlying lattice model, where the excitation has non-linear dispersion. We will discuss how to adapt eq.~(\ref{Enmiua}) and eq.~(\ref{Enmiud}) to describe the dynamics of charged R\'enyi (logarithmic) negativity in the mass quench protocol of the complex harmonic chain.
\section{Charge imbalance resolved negativity}\label{section5}
In this section, using the results obtained in the last section, we will compute the dynamics of the charge imbalance resolved negativity between two intervals after a global quench to a conformal Hamiltonian. According to the discussion in section \ref{section2}, our strategy is to first compute $\mZ_{T_2}(\mathrm{q})$ and $p(\mathrm{q})$ (cf. eq.~(\ref{ZT2q}) and eq.~(\ref{pq})) from the charged moments of partially transposed RDM, and then the charge imbalance resolved negativity is given by eq.~(\ref{ZT2q}).
\subsection{Bipartite system}
$\pmb{t<l_1/2}$\\
\\
In this time region, we have
\be
\mathcal{E}(\m)=-h(\m)\frac{\pi t}{\tau_0},\qquad \mathcal{E}_1(\m)=-h_1(2\m)\frac{\pi t}{\tau_0}.
\ee
After Fourier transformation, we find
\be
\mathcal{Z}_{T_2}(\mathrm{q})=\frac{(-1)^{\mathrm{q}} e^{\frac{\pi \mathrm{q}^2}{4t/\tau_0}-\frac{\pi t}{2\tau_0}}}{2\sqrt{t/\tau_0}}
\Re\left[\mathrm{Erfi}\left(\frac{\sqrt{\pi}(i\mathrm{q}+2t/\tau_0)}{2\sqrt{t/\tau_0}}\right)\right],
\ee
where $\mathrm{Erfi}(x)$ is the imaginary error function
\be\label{Erfi}
\mathrm{Erfi}(x)=\frac{-2i}{\sqrt{\pi}}\int_0^{ix}dte^{-t^2}\xrightarrow{x\rightarrow\infty}\frac{e^{x^2}}{\sqrt{\pi}x}.
\ee
A very similar expression exists for $p(\mathrm{q})$ which we omit here. In the spacetime scaling limit, we have much more concise results
\be
\mathcal{Z}_{T_2}(\mathrm{q})=\frac{2e^{\frac{\pi t}{2\tau_0}}t/\tau_0}{\pi \mathrm{q}^2+4\pi t^2/\tau_0^2}
\ee
and the probability distribution is given by
\be
p(\mathrm{q})=\frac{(1+(-1)^{\mathrm{q}})t/\tau_0}{\pi \mathrm{q}^2+\pi t^2/\tau_0^2}.
\ee
Thus the charge imbalance resolved negativity is obtained from eq.~(\ref{ZT2q})
\be
\mathcal{N}(\mathrm{q})=\frac{e^{\frac{\pi t}{2\tau_0}}(\mathrm{q}^2+t^2/\tau_0^2)}{(1+(-1)^{\mathrm{q}})(\mathrm{q}^2+4 t^2/\tau_0^2)}-\frac12.
\ee
\\
$\pmb{t>l_1/2}$\\
\\
In this time region, we can simply make the replacement $t\rightarrow l_1/2$ to obtain corresponding quantities. In particular, the charge imbalance resolved negativity is given by
\be
\mathcal{N}(\mathrm{q})=\frac{e^{\frac{\pi l_1}{4\tau_0}}(4\mathrm{q}^2+l_1^2/\tau_0^2)}{4(1+(-1)^{\mathrm{q}})(\mathrm{q}^2+l_1^2/\tau_0^2)}-\frac12.
\ee
\subsection{Two adjacent intervals}
Without loss of generality, we will assume $l_1<l_2$, the other case can be worked out similarly.\\
\\
$\pmb{t<l_1/2<l_2/2<(l_1+l_2)/2}$\\
\\
For this early time, we have
\be
\mathcal{E}(\m)=-(2h(\m)-h_2(\m))\frac{\pi t}{2\tau_0},\qquad \mathcal{E}_1(\m)=-(2h_1(2\m)-h_3(\m))\frac{\pi t}{2\tau_0}.
\ee
Then after Fourier transformation, we find
\be
\mathcal{Z}_{T_2}(\mathrm{q})=\frac{(-1)^{\mathrm{q}}e^{\frac{\pi \mathrm{q}^2}{3t/\tau_0}-\frac{\pi t}{2}}}{\sqrt{3t/\tau_0}}
\Re\left[\mathrm{Erfi}\left(\frac{\sqrt{\frac{\pi}{3}}(2i\mathrm{q}+3t/\tau_0)}{2\sqrt{t/\tau_0}}\right)\right],
\ee
Therefore in the spacetime scaling limit ($t\gg\tau_0$), we have
\be
\mathcal{Z}_{T_2}(\mathrm{q})=\frac{6e^{\frac{\pi t}{4\tau_0}}t/\tau_0}{4\pi \mathrm{q}^2+9\pi(t/\tau_0)^2}.
\ee
In the spacetime scaling regime, the probability distribution can be computed similarly and the finial result is
\be
p(\mathrm{q})=\frac{t/\tau_0}{\pi \mathrm{q}^2+\pi t^2/\tau_0^2}.
\ee
Then the charge imbalance resolved negativity is given by
\be
\mathcal{N}(\mathrm{q})=\frac{3e^{\frac{\pi t}{4\tau_0}}(\mathrm{q}^2+t^2/\tau_0^2)}{4\mathrm{q}^2+9t^2/\tau_0^2}-\frac{1}{2}.
\ee
\\
$\pmb{l_1/2<t<l_2/2<(l_1+l_2)/2}$\\
\\
In this case we have
\be
\mathcal{E}(\m)=-(h(\m)-h_2(\m))\frac{\pi t}{2\tau_0}-h(\m)\frac{\pi l_1}{4\tau_0}.
\ee
Then
\be
\mathcal{Z}_{T_2}(\mathrm{q})=\frac{(-1)^{\mathrm{q}}e^{\frac{\pi \mathrm{q}^2}{(l_1+t)/\tau_0}-\frac{\pi(l_1+2t)}{8\tau_0}}}{\sqrt{(l_1+t)/\tau_0}}
\Re\left[\mathrm{Erfi}\left(\frac{\sqrt{\pi}(2i\mathrm{q}+(l_1+t)/\tau_0)}{2\sqrt{(l_1+t)/\tau_0}}\right)\right].
\ee
In the spacetime scaling limit
\be
\mathcal{Z}_{T_2}(\mathrm{q})=\frac{2e^{\frac{\pi l_1}{8\tau_0}}(l_1+t)/\tau_0}{4\pi \mathrm{q}^2+\pi(l_1+t)^2/\tau_0^2}.
\ee
The probability distribution is
\be
p(\mathrm{q})=\frac{4(l_1+2t)/\tau_0}{16\pi \mathrm{q}^2+\pi(l_1+2t)^2/\tau_0^2}.
\ee
Then the charge-imbalance resolved negativity is
\be
\mathcal{N}(\mathrm{q})=\frac{16\mathrm{q}^2\tau_0^2+(l_1+2t)^2}{16\mathrm{q}^2\tau_0^2+4(l_1+t)^2}\frac{(l_1+t)e^{\frac{\pi l_1}{8\tau_0}}}{l_1+2t}-\frac{1}{2}.
\ee
\\
$\pmb{l_2/2<t<(l_1+l_2)/2}$\\
\\
In this time region, we have
\be
\mathcal{E}(\m)=h_2(\m)\frac{\pi t}{2\tau_0}-h(\m)\frac{(l_1+l_2)\pi}{4\tau_0},\qquad\mathcal{E}_1(\m)=-\frac{\pi}{4\tau_0}(h_1(2\m)(l_1+l_2)-2th_3(\m)).
\ee
Then applying Fourier transformation, we find
\be
\mathcal{Z}_{T_2}(\mathrm{q})=\frac{(-1)^{\mathrm{q}}e^{\frac{\pi \mathrm{q}^2}{(l-t)/\tau_0}-\frac{\pi l}{8\tau_0}}}{\sqrt{(l-t)/\tau_0}}
\Re\left[\mathrm{Erfi}\left(\frac{\sqrt{\pi}(2i\mathrm{q}+(l-t)/\tau_0)}{2\sqrt{(l-t)/\tau_0}}\right)\right].
\ee
Here and in the following sections we will always use the definition $l\equiv l_1+l_2$. In the spacetime scaling limit
\be
\mathcal{Z}_{T_2}(\mathrm{q})=\frac{2e^{\frac{\pi(l-2t)}{8\tau_0}}(l-t)/\tau_0}{4\pi \mathrm{q}^2+\pi(l-t)^2/\tau_0^2}.
\ee
the probability distribution is
\be
p(\mathrm{q})=\frac{4l/\tau_0}{16\pi \mathrm{q}^2+\pi l^2/\tau_0^2}.
\ee
the charge imbalance resolved negativity is then easily obtained as
\be
\mathcal{N}(\mathrm{q})=\frac{16\mathrm{q}^2\tau_0^2+l^2}{16\mathrm{q}^2\tau_0^2+4(l-t)^2}\frac{(l-t)e^{\frac{\pi(l-2t)}{8\tau_0}}}{l}-\frac{1}{2}.
\ee
\\
$\pmb{t>(l_1+l_2)/2}$\\
\\
In this case, we have
\be
\begin{split}
\mathcal{E}(\m)=-[h(\m)-h_2(\m)]\frac{\pi(l_1+l_2)}{4\tau_0}=\mathcal{E}_1(\m).
\end{split}
\ee
Then the charge imbalance resolved negativity is
\be
\mathcal{N}(\mathrm{q})=0.
\ee
\subsection{Two disjoint intervals}
The computation of charge imbalance resolved negativity for two disjoint intervals is similar to previous subsection, here we just report the final results (the space time scaling limit has been taken). In this circumstance, the different size relation between $l_1,l_2$ and $d$ may lead to different behavior. However, for simplicity, we will only consider the case $d<l_1<l_2$.\\
\\
$\pmb{0<t<d/2<l_1/2}$\\
\\
At this early time stage, we have $\mathcal{E}(\m)=\mathcal{E}_1(\m)$, therefore
\be
\mathcal{N}(\mathrm{q})=0.
\ee
\\
$\pmb{d/2<t<l_1/2}$\\
\\
In this time period, we have
\be
\mathcal{Z}_{T_2}(\mathrm{q})=\frac{4e^{\frac{\pi(2t-d)}{8\tau_0}}(6t-d)/\tau_0}{16\pi \mathrm{q}^2+\pi (6t-d)^2/\tau_0^2},\quad p(\mathrm{q})=\frac{t/\tau_0}{\pi \mathrm{q}^2+\pi t^2/\tau_0^2}.
\ee
Then the charge imbalance resolved negativity is obtained straightforward as
\be
\mathcal{N}(\mathrm{q})=\frac{2e^{\frac{\pi(2t-d)}{8\tau_0}}(\mathrm{q}^2\tau_0^2+t^2)}{16\mathrm{q}^2\tau_0^2+(6t-d)^2}\frac{6t-d}{t}-\frac12.
\ee
\\
$\pmb{l_1/2<l_2/2<t<(l_1+d)/2}$\\
\\
In this time region, the following results are obtained (in the spacetime scaling limit)
\be
\mathcal{Z}_{T_2}(\mathrm{q})=\frac{4e^{\frac{\pi(2t-d)}{8\tau_0}}(l-d+2t)/\tau_0}{16\pi \mathrm{q}^2+\pi(l-d+2t)^2/\tau_0^2},\quad p(\mathrm{q})=\frac{4l/\tau_0}{16\pi \mathrm{q}^2+\pi l^2/\tau_0^2}.
\ee
The charge imbalance resolved negativity is given by
\be
\mathcal{N}(\mathrm{q})=\frac{e^{\frac{\pi(2t-d)}{8\tau_0}}(16\mathrm{q}^2\tau_0^2+l^2)}{32\mathrm{q}^2\tau_0^2+2(2t+l-d)^2}\frac{2t+l-d}{l}-\frac12.
\ee
\\
$\pmb{(l_2+d)/2<t<(l_1+l_2+d)/2}$\\
\\
In this time period, we have
\be
\mathcal{Z}_{T_2}(\mathrm{q})=\frac{4e^{\frac{\pi(d+l-2t)}{8\tau_0}}(d+2l-2t)/\tau_0}{16\pi\mathrm{q}^2+\pi(d+2l-2t)^2/\tau_0^2},\quad p(\mathrm{q})=\frac{4l/\tau_0}{16\pi \mathrm{q}^2+\pi l^2/\tau_0^2}.
\ee
The charge imbalance resolved negativity is
\be
\mathcal{N}(\mathrm{q})=\frac{e^{\frac{\pi(d+l-2t)}{8\tau_0}}(16\mathrm{q}^2\tau_0^2+l^2)}{32\mathrm{q}^2\tau_0^2+2(d+2l-2t)^2}\frac{d+2l-2t}{l}-\frac12.
\ee
\\
$\pmb{t>(l_1+l_2+d)/2}$\\
\\
In this late time region, we have $\mathcal{E}(\m)=\mathcal{E}_1(\m)$ again, and charge imbalance resolved negativity is zero
\be
\mathcal{N}(\mathrm{q})=0.
\ee
\begin{figure}
        \centering
        \subfloat
        {\includegraphics[width=7cm]{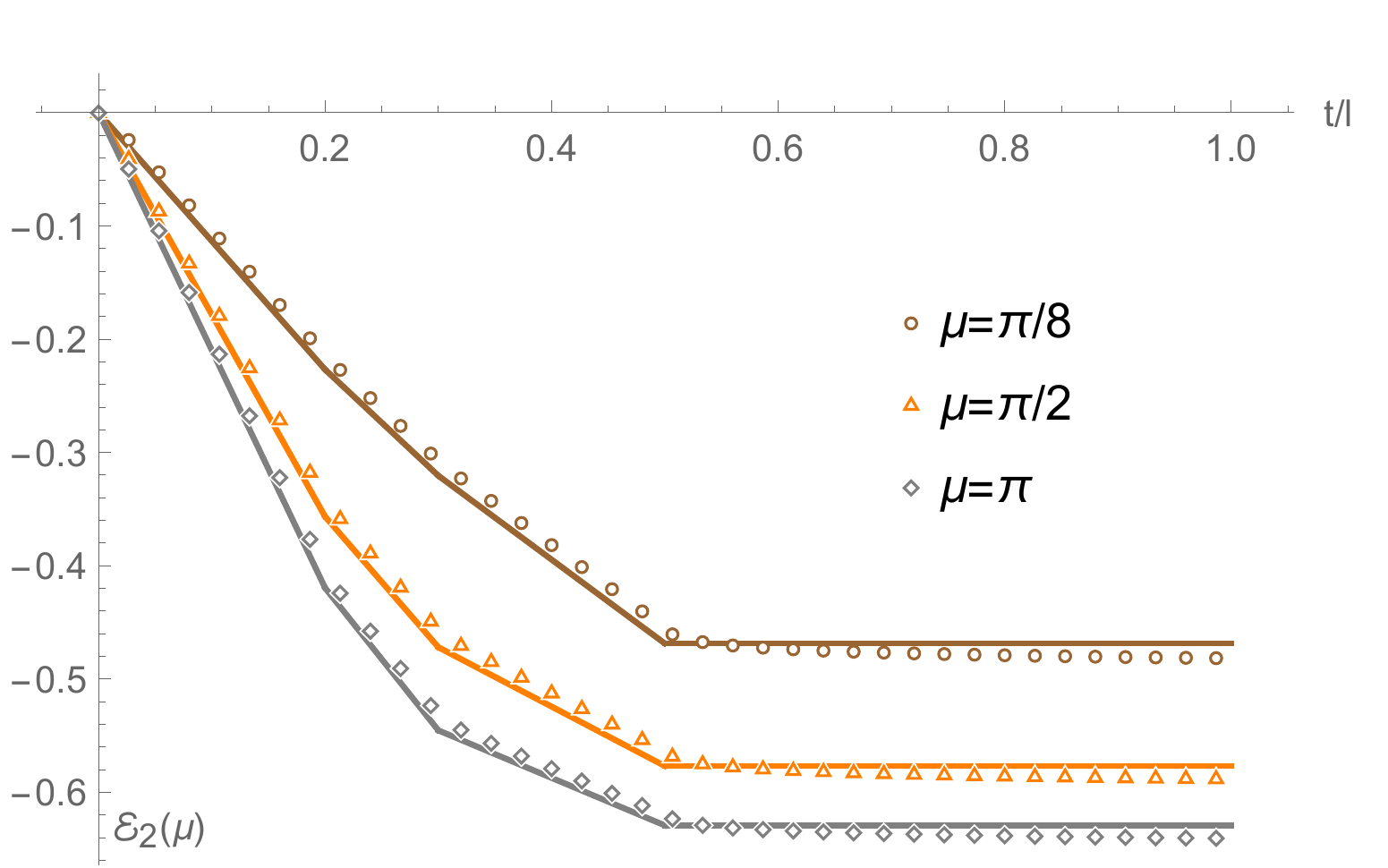}} \quad\quad
        {\includegraphics[width=7cm]{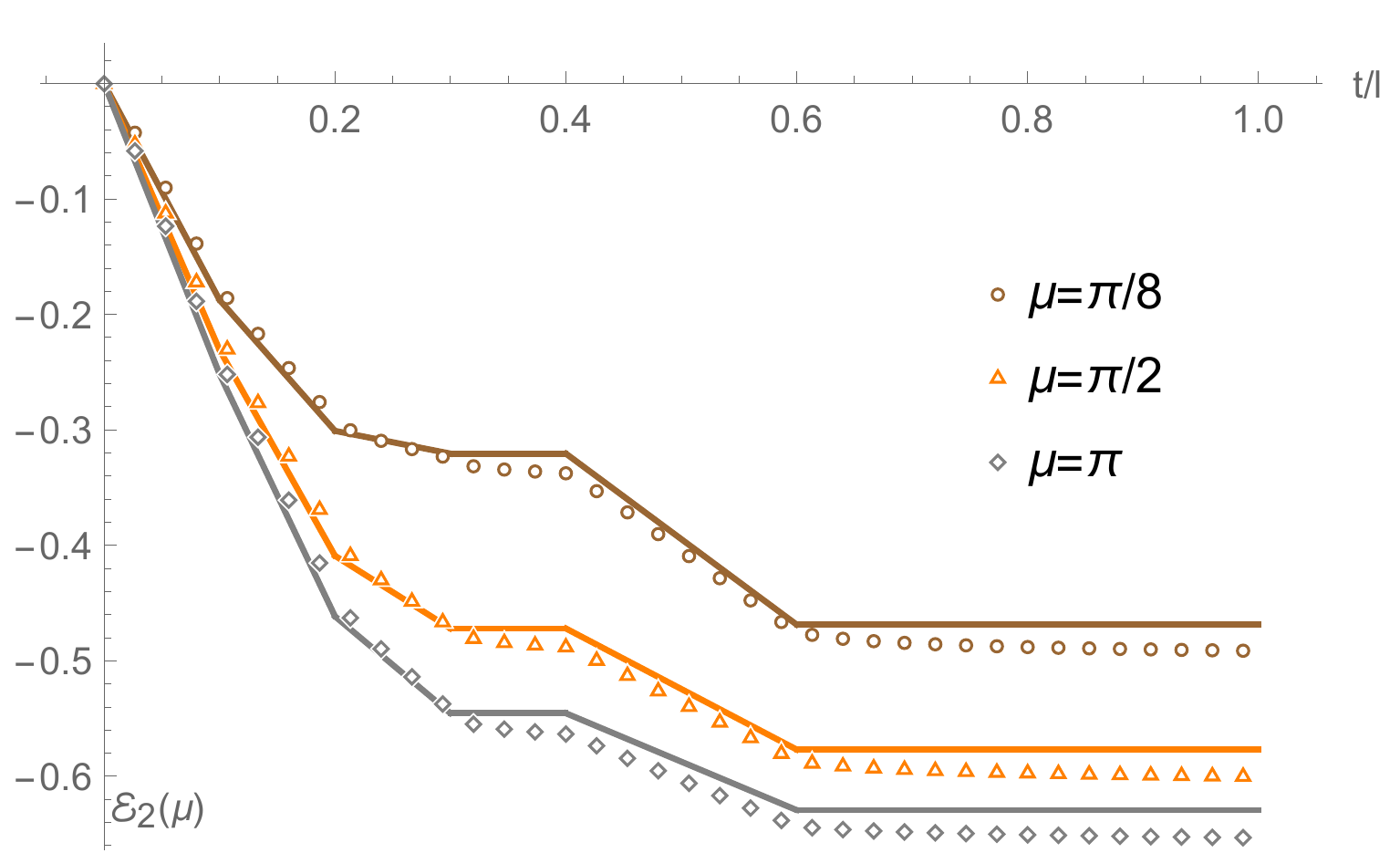}}
        \caption{Numerical data of charged R\'enyi logarithmic negativity $\mathcal{E}_2(\m)$ as a function of $t/l$ for $\m=\frac{\pi}{8},\frac{\pi}{2},\pi$ in the complex harmonic chain. The full lines are the CFT predictions (cf. eq.~(\ref{Enmiua}) and eq.~(\ref{Enmiud})). Left panel: Adjacent interval with $L=3000,l_1=200,l_2=300$. Right panel: Disjoint interval with $L=3000, l_1=200,l_2=300,d=100$. We have used the best fitted value of $\tau_0$.}
        \label{fig1}
\end{figure}
\subsection{Total negativity}
Having obtained the quench dynamics of the charge imbalance resolved negativity, it's an easy task to check whether we can recover the known results of quench dynamics of entanglement negativity. Using the formula
\be
\sum_{\mathrm{q}=-\infty}^{+\infty}\frac{1}{\pi}\frac{a^2}{\mathrm{q}^2+a^2}=\coth(\pi a)\xrightarrow{a\rightarrow\infty}1
\ee
and the reconstruction formula of negativity eq.~(\ref{Recon}), it's easy to check that indeed in each time period, we can obtain the total entanglement negativity from the charge imbalance resolved one, i.e. in the spacetime scaling limit, we have the following relation
\be
\sum_{\mathrm{q}=-\infty}^{+\infty}p(\mathrm{q})\mathcal{N}(\mathrm{q})=\frac12\exp\{\mathcal{E}(\m=0)\}-\frac12=\mathcal{N}.
\ee
In this way, we have recovered the known quench dynamics \cite{Coser:2014gsa} of the total logarithmic negativity from the charge imbalance resolved ones, providing evidence of the correctness of our results.
\section{Numerical test}\label{section6}
In this section, we test our analytical predictions against numerical computation in the complex harmonic chain which is the lattice version of our complex Klein-Gordon field theory. We will use the correlation matrix techniques to obtain the charged R\'enyi (logarithmic) negativity.
\par In section \ref{section3}, we have already reported the Hamiltonian and the $U(1)$ conserved charge of the complex harmonic chain (cf. eq.~(\ref{HCHC1}) and eq.~(\ref{Q})). From the dispersion relation, it's easy to see that the Hamiltonian has zero modes for $k=0$ and $m=0$, and the group velocities are obtained from the dispersion relation as
\be
v_p\equiv\frac{\p e(p)}{\p p}=\frac{\sin(p)}{\sqrt{m^2+4\sin^2(p/2)}},\qquad p\equiv\frac{2\pi k}{L}.
\ee
The maximum velocity $v_{\text{max}}\equiv \max_pv_p$ determine s the spreading of entanglement and correlations.
\par In the global quench protocol, for $t<0$, the system is prepared in the ground state of the Hamiltonian
\be
H_{CHC}(m_0)=\sum_{k=0}^{L-1}e_{0,k}(a_k^{\dg}a_k+b_k^{\dg}b_k),\quad e_{0,k}=\sqrt{m_0^2+4\sin^2\lt(\frac{\pi k}{L}\rt)}.
\ee
At $t=0$, the system evolves unitarily according to a new Hamiltonian $H_{CHC}(m)$ with a different value $m$, namely $\ket{\psi(t)}=e^{-iH_{CHC}(m)t}\ket{\psi_0}$. The dynamics of symmetry resolved negativity after a global quench described in the previous section corresponds to set the new parameter $m=0$. In this section, we will focus on the global quench to a massless Hamiltonian (i.e. $m=0$) to test our CFT predictions. In the next section we will discuss the case $m\neq 0$.
\par Since the Hamiltonian of a complex harmonic chain is just two copies of a real harmonic chain's Hamiltonian, we could focus on the real harmonic chain first, and take the double copies into account only at the end of the calculations. In the correlation matrix method of computing entanglement measures, we first need to know the following two-point correlators of the real scalars
\be
\begin{split}
\mathbf{X}_{rs}(t)\equiv\bra{\psi_0}\phi_r(t)\phi_s(t)\ket{\psi_0},\\
\mathbf{P}_{rs}(t)\equiv\bra{\psi_0}\pi_r(t)\pi_s(t)\ket{\psi_0},\\
\mathbf{M}_{rs}(t)\equiv\bra{\psi_0}\phi_r(t)\pi_s(t)\ket{\psi_0},\\
\end{split}
\ee
where
\be
\phi_r(t)=e^{iH_{HC}(m)t}\phi_r(0)e^{-iH_{HC}(m)t},\qquad \pi_r(t)=e^{iH_{HC}(m)t}\pi_r(0)e^{-iH_{HC}(m)t}.
\ee
The explicit form of these correlators is given by
\be
\begin{split}
&\mathbf{X}_{rs}(t)=\frac{1}{2L}\sum_{k=0}^{L-1}X_k(t)\cos\left[\frac{2\pi k}{L}(r-s)\right],\\
&\mathbf{P}_{rs}(t)=\frac{1}{2L}\sum_{k=0}^{L-1}P_k(t)\cos\left[\frac{2\pi k}{L}(r-s)\right],\\
&\mathbf{M}_{rs}(t)=\frac{i}{2}\delta_{rs}-\frac{1}{2L}\sum_{k=0}^{L-1}M_k(t)\cos\left[\frac{2\pi k}{L}(r-s)\right],
\end{split}
\ee
where
\be
\begin{split}
&X_k(t)=\frac{1}{e_k}\left(\frac{e_k}{e_{0,k}}\cos^2(e_kt)+\frac{e_{0,k}}{e_{k}}\sin^2(e_kt)\right),\\
&P_k(t)=e_k\left(\frac{e_k}{e_{0,k}}\sin^2(e_kt)+\frac{e_{0,k}}{e_{k}}\cos^2(e_kt)\right),\\
&M_k(t)=\left(\frac{e_k}{e_{0,k}}-\frac{e_{0,k}}{e_{k}}\right)\sin(e_kt)\cos(e_kt).
\end{split}
\ee
\begin{figure}
        \centering
        \subfloat
        {\includegraphics[width=7cm]{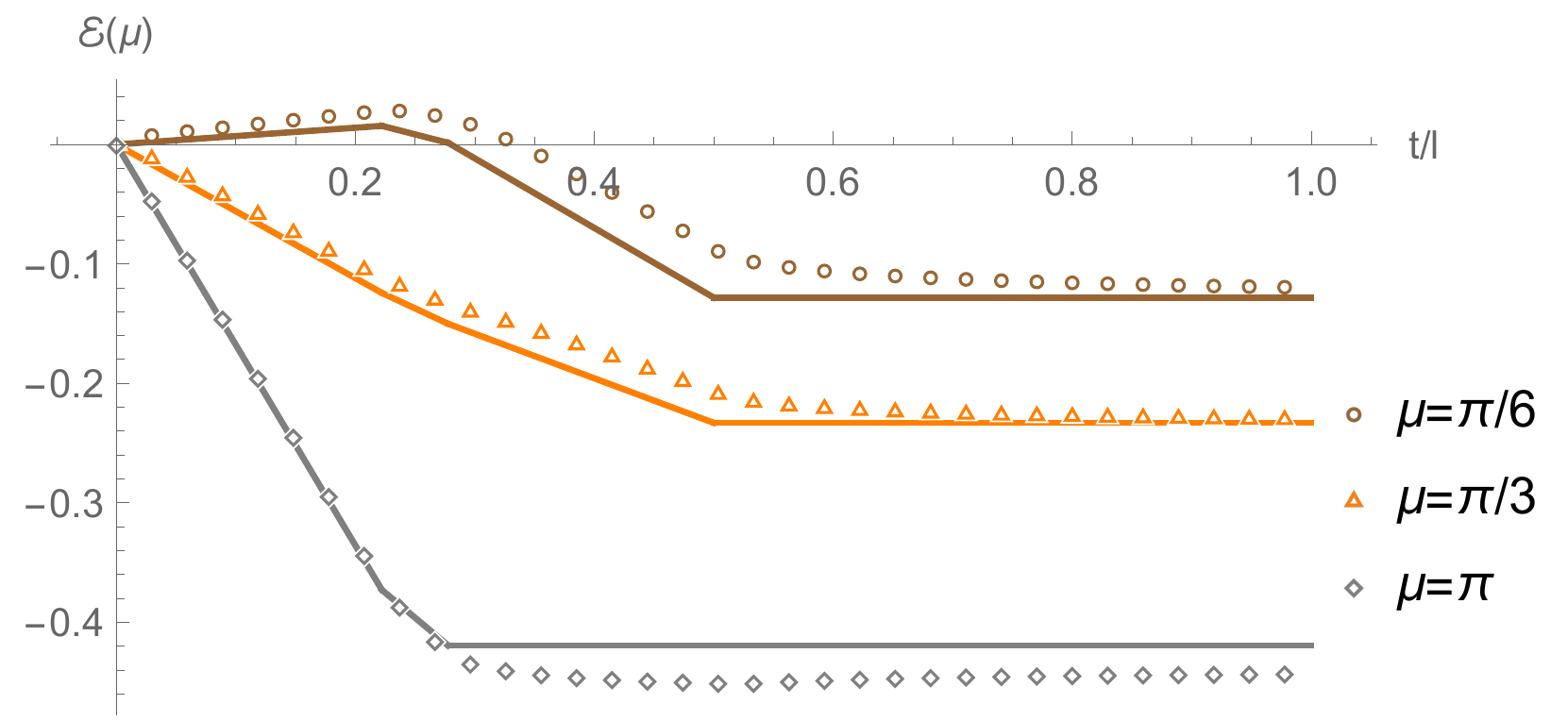}} \quad\quad
        {\includegraphics[width=7cm]{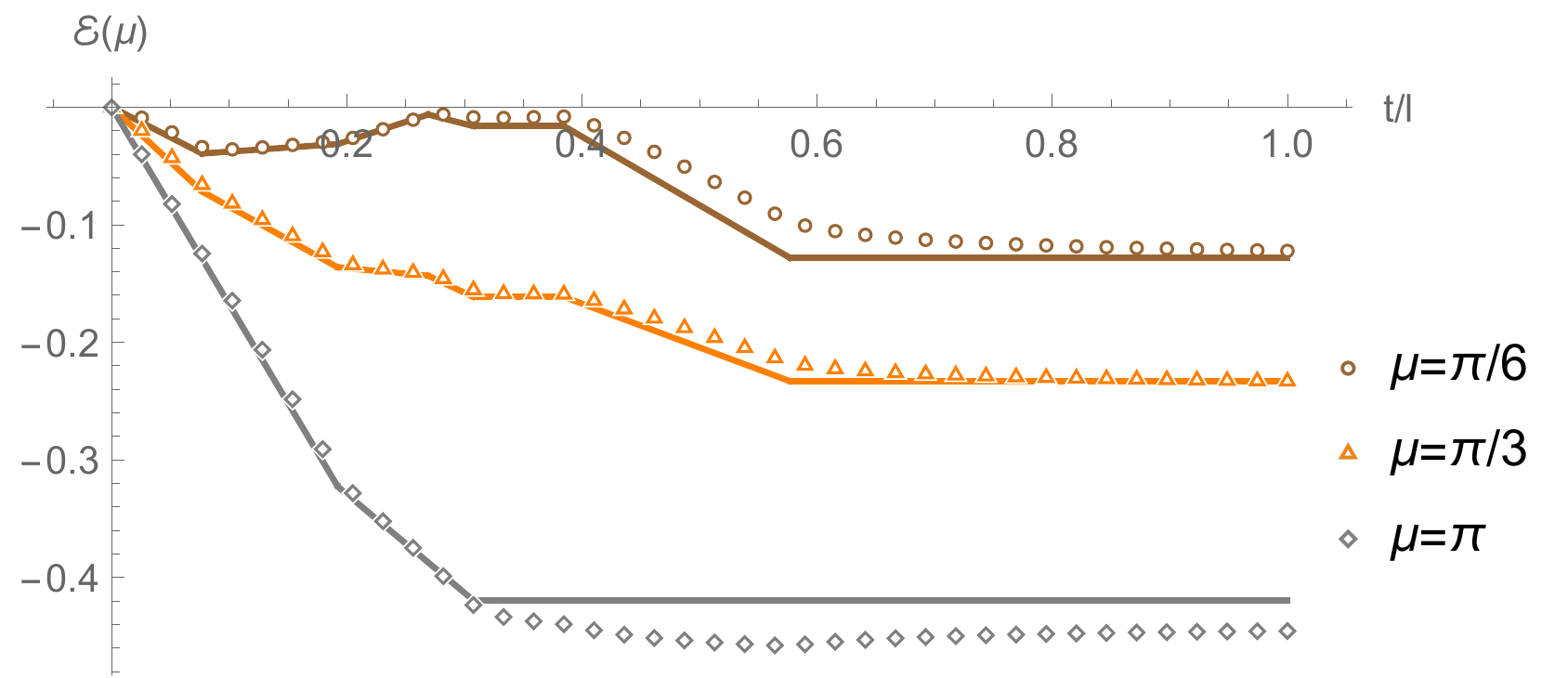}}
        \caption{Numerical data of charged logarithmic negativity $\mathcal{E}(\m)$ as a function of $t/l$ for $\m=\frac{\pi}{6},\frac{\pi}{3},\pi$ in the complex harmonic chain. The full lines are the CFT predictions (cf. eq.~(\ref{Emiua}) and eq.~(\ref{Emiud})). Left panel: Adjacent interval with $L=2000,l_1=200,l_2=250$. Right panel: Disjoint interval with $L=5000, l_1=250,l_2=400,d=100$. We have used the best fitted value of $\tau_0$.}
        \label{fig2}
\end{figure}
From the above expressions, it's easy to find that the contribution from the zero mode $m=0$ and $k=0$ is finite
\be
X_0(t)=m_0^{-1}+m_0t^2,\quad P_0(t)=m_0,\quad M_0(t)=-m_0t.
\ee
\par The evolution of charged R\'enyi (logarithmic) negativity of any subsystem $A$ containing $l$ lattice sites can be computed from these time-dependent correlators. Firstly, we should consider the correlation matrices $\mathbf{X}_A(t)$,$\mathbf{P}_A(t)$ and $\mathbf{M}_A(t)$ obtained by restricting the the indices of the corresponding correlation matrices to the sites belonging to $A$. Given $\mathbf{X}_A$,$\mathbf{P}_A$ and $\mathbf{M}_A$, the covariance matrix $\G_A$ and the symplectic matrix $J_A $ associated to the subsystem $A$ are
\be
\G_A(t)=\begin{pmatrix}
\mathbf{X}_A(t)&\mathbf{M}_A(t)\\
\mathbf{M}_A(t)^{\mathrm{t}}&\mathbf{P}_A(t)
\end{pmatrix},\qquad
J_A=\begin{pmatrix}
\mathbf{0}_l&\mathbf{I}_l\\
-\mathbf{I}_l&\mathbf{0}_l
\end{pmatrix},\qquad
\ee
where $\mathbf{0}_l$ and $\mathbf{I}_l$ are $l\times l$ zero matrix and identity matrix respectively.
Then find the eigenvalues of the $2l\times 2l$ matrix $iJ_A\G_A(t)$ which we denoted it by $\{\pm\s_1(t),\cdots,\pm\s_l(t)\}$. It's also convenient to introduce the Fock space basis $\ket{\mathbf{n}}\equiv\otimes_{j=1}^l\ket{n_j}$, defined by products of eigenstates of the number operator in the subsystem $A$, the reduced density matrix of $A$ (in a real harmonic chain) can be written as
\be
\rho_A(t)=\sum_{\mathbf{n}}\prod_{j=1}^{l}\frac{1}{\s_j(t)+1/2}\lt(\frac{\s_j(t)-1/2}{\s_j(t)+1/2}\rt)^{n_j}\ket{\mathbf{n}}\bra{\mathbf{n}}.
\ee
\par In the Fock basis $\{\ket{\mathbf{n}}\}$, $Q_2^{T_2}=Q_2$ and the operator $\mQ_A=Q_1-Q^{T_2}_2=Q_1-Q_2$ becomes exactly the charge imbalance operator. For the complex harmonic chain, the charged R\'enyi negativity factorises as
\be
R_{n}(\m)=\Tr[(\rho^{T_2}_A)^ne^{\mathrm{i}\m \mQ_A^a}]\times\Tr[(\rho^{T_2}_A)^ne^{-\mathrm{i}\m \mQ_A^b}].
\ee
\par In bosonic system, the net effect of partial transposition with respect to $A_2$ is changing the sign of the momenta corresponding to $A_2$. Thus the momenta correlators in the partial transposed density matrix can be obtained from $\mathbf{P}_A$ by simply change the sign of the momenta that in $A_2$, i.e. $\mathbf{P}_A^{T_2}=\mathbf{R}_{A_2}\mathbf{P}_A\mathbf{R}_{A_2}$,where $\mathbf{R}_{A_2}$ is the $l_2\times l_2$ diagonal matrix with elements $(\mathbf{R}_{A_2})_{rs}=(-1)^{\d_{r\in A_2}}\d_{rs}$. Thus the partial transformed covariance matrix is
\be
\G^{T_2}_A(t)=\begin{pmatrix}
\mathbf{I}_l&\mathbf{0}_l\\
\mathbf{0}_l&\mathbf{R}_{A_2}
\end{pmatrix}\G_A(t)\begin{pmatrix}
\mathbf{I}_l&\mathbf{0}_l\\
\mathbf{0}_l&\mathbf{R}_{A_2}.
\end{pmatrix}
\ee
If we denote the eigenvalues of $iJ_A\G_A(t)^{T_2}$ by $\{\pm\tau_1(t),\pm\tau_2(t),\cdots,\pm\tau_l(t)\}$, then the charged R\'enyi logarithmic negativity is given by\footnote{This formula is consistent with the eq.(6.18) in paper \cite{Chen:2021nma}.}
\be
\mathcal{E}_n(\m)=-2\sum_{j=1}^l\log\Big|\lt(\t_j(t)+\frac12\rt)^n-e^{\mathrm{i}\m}\lt(\t_j(t)-\frac12\rt)^n\Big|
\ee
and the charged logarithmic negativity is
\be
\mathcal{E}(\m)=-2\sum_{j=1}^l\log\Big||\t_j(t)+\frac12|-e^{\mathrm{i}\m}|\t_j(t)-\frac12|\Big|.
\ee
\par The numerical data of the dynamics of the charged R\'enyi (logarithmic) negativity are shown in fig.~\ref{fig1} and fig.~\ref{fig2}, in which the CFT predictions are drawn with the full line for comparison.
\section{Global mass quench of the harmonic chain}\label{section7}
In this section, we consider a global quantum quench in which the complex harmonic chain is initially prepared in the ground state with mass $m_0$ and at time $t=0$ the mass is quenched to a different (non-zero) value $m\neq m_0$.
\subsection{Quasi-particle picture for entanglement entropy}
In free scalar theory, since the particle numbers in each momentum mode are conserved quantities, the time evolution leads to local and quasi-local observables converging to their average values in the Generalized Gibbs Ensemble (GGE) instead of ordinary Gibbs ensemble.
\par The quasi-particle picture of the time evolution of entanglement after a global quantum quench has been proposed in \cite{Calabrese:2005in}, The basic idea is that one can view the initial state as a source of quasi-particle excitations. Pairs of particles emitted from the same point in space are highly entangled whereas particles produced from points far apart are incoherent. We assume that the pairs of quasi-particle are created uniformly with opposite momenta $(p,-p)$. After the production, these quasi-particles travel ballistically with velocity $v_p=-v_{-p}$. The entanglement entropy and R\'enyi entropy of subsystem $A$ are proportional to the pairs of entangled quasi-particle shared with its complement at a given time $t$.
\par In free models, we have
\be\label{Snlat}
S_n(t)=\int\frac{dp}{2\pi}s^{(n)}_{GGE}(p)\min(2|v_p|t,l),
\ee
with $l$ being the length of subsystem $A$ and $s^{(n)}_{GGE}(p)$ is momentum space density of the R\'enyi entropies in the GGE thermodynamic state \cite{vidmar2016generalized}. Recall that the well-known result of the CFT prediction of the quench to massless dynamics of R\'enyi entropies
\be\label{Sncft}
S_n(t)=\frac{1}{n-1}\frac{\pi\D_{n,0}}{2\tau_0}\min(2t,l).
\ee
We can formally obtained the formula eq.~(\ref{Snlat}) from the CFT prediction eq.~(\ref{Sncft}) by replacing $t\rightarrow|v_p|t$, $-\frac{\pi\D_{n,\m}}{2\tau_0}\rightarrow s^{(n)}_{GGE}(p)$ and then integrating over all possible $p$.
\begin{figure}
        \centering
        \subfloat
        {\includegraphics[width=7cm]{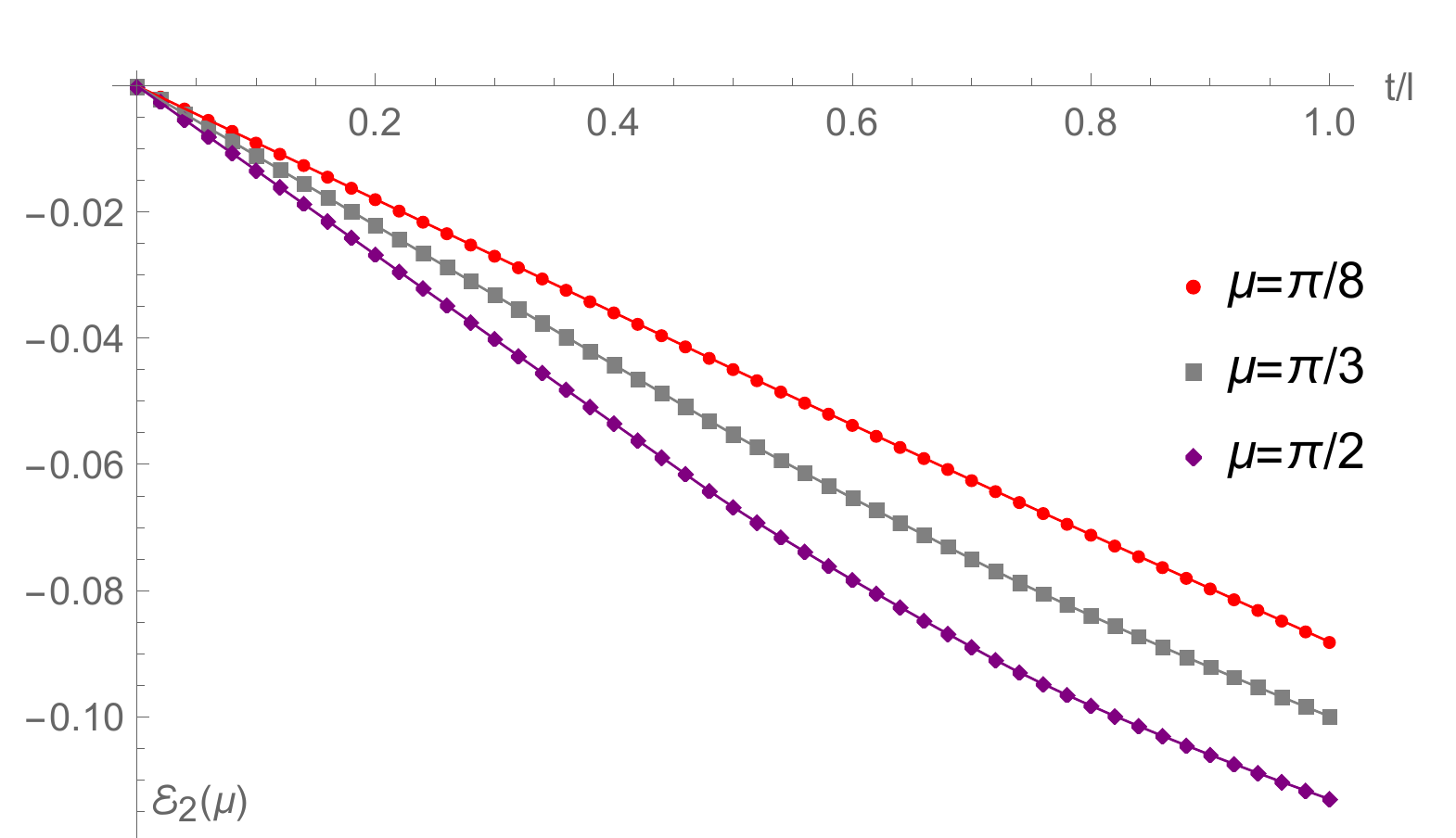}} \quad\quad
        {\includegraphics[width=7cm]{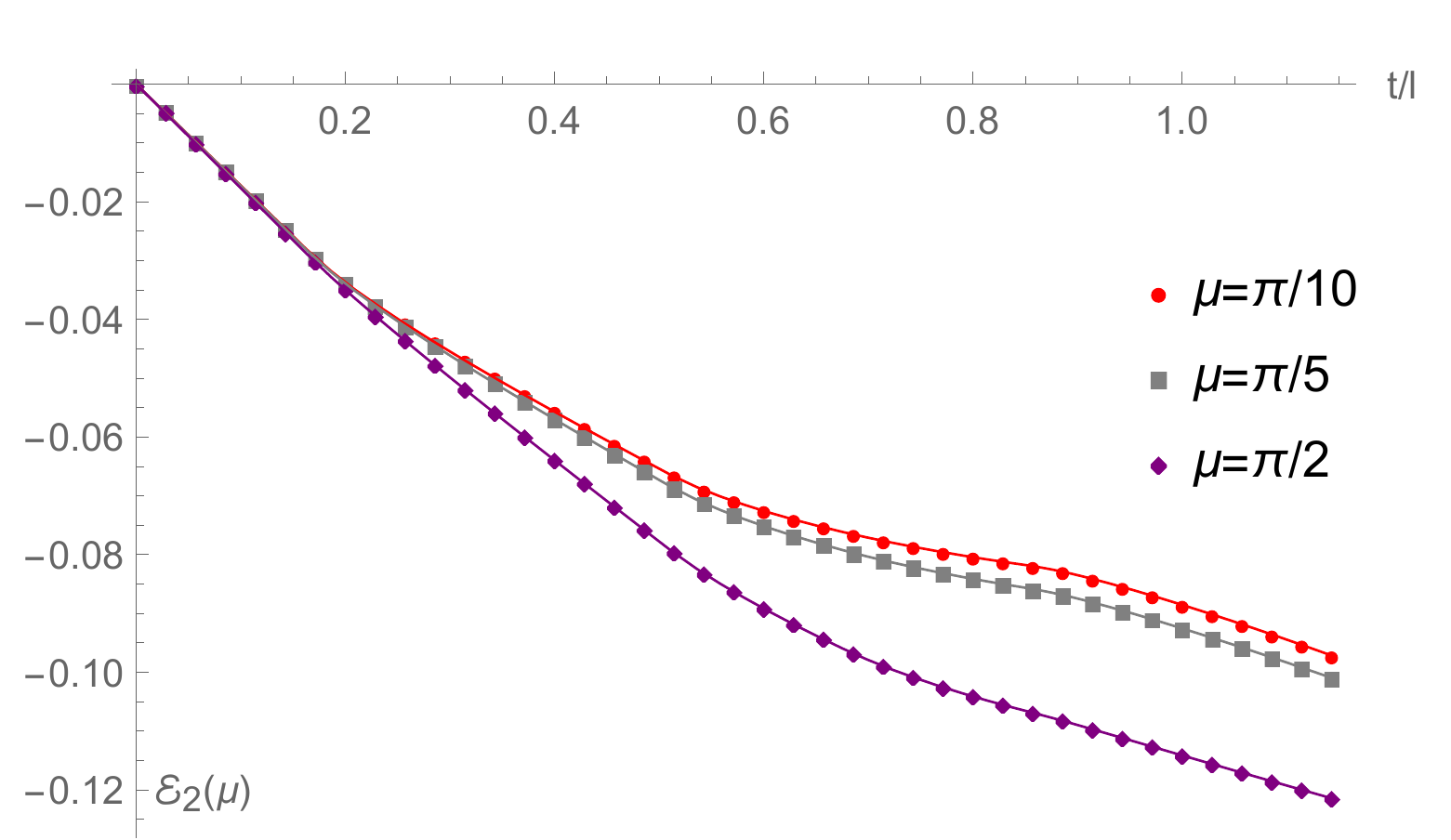}}
        \caption{Charged R\'enyi logarithmic negativity $\mathcal{E}_2(\m)$ as a function of $t/l$ for different $\m$ after the mass quench from $m_0=1$ to $m=2$ in the complex harmonic chain. The full lines are the quasi-particle predictions (cf. eq.~(\ref{EnmiuMa}) and eq.~(\ref{EnmiuMd})). Left panel: Adjacent intervals with $L=2500,l_1=200,l_2=300$ and for $\m=\frac{\pi}{8},\frac{\pi}{3},\frac{\pi}{2}$. Right panel: Disjoint intervals with $L=2500,l_1=150,l_2=200,d=50$ and for $\m=\frac{\pi}{10},\frac{\pi}{5},\frac{\pi}{2}$. As shown in the figure, the agreement is extremely excellent.}
        \label{fig3}
\end{figure}
\subsection{Quasi-particle picture for charged logarithmic negativity}
\par The complex harmonic chain is a free models and the stationary behavior of local and quasi-local observables is described by the GGE
\be
\rho_{GGE}=Z^{-1}e^{-\sum_k\l_k^{(a)}a_k^{\dg}a_k}\otimes e^{-\sum_k\l_k^{(b)}b_k^{\dg}b_k},
\ee
where $\l_k^{(a)},\l_k^{(b)}$ are Lagrange mulitpliers and $Z$ is the normalisation constant such that $\Tr\rho_{GGE}=1$. We have
\be
Z=\prod_{i=a,b}\Tr e^{-\sum_k\l_k^{(i)}\hat{n}_k^{(i)}}=\prod_{i=a,b}\prod_k\sum_{n_k^{(i)}=0}^{\inf}e^{-\l_k^{(i)}n_k^{(i)}}=\prod_{i=a,b}\prod_k(1-e^{-\l_k^{(i)}})^{-1}.
\ee
The mode occupation number $\hat{n}_k^{(a)}\equiv a_k^{\dg}a_k,\hat{n}_k^{(b)}\equiv b_k^{\dg}b_k$ are conserved quantities. Therefore we have
\be
\bra{\psi_0}\hat{n}_{k}^{(i)}\ket{\psi_0}=\Tr[\hat{n}_{k}^{(i)}\rho_{GGE}]=(e^{\l_k^{(i)}}-1)^{-1}.
\ee
In the ground state $\ket{\psi_0}$, we have  $\bra{\psi_0}\hat{n}_{k}^{(a)}\ket{\psi_0}=\bra{\psi_0}\hat{n}_{k}^{(b)}\ket{\psi_0}\equiv n_k$, thus
\be
\l_k^{(a)}=\l_k^{(b)}=\log(1+n_k^{-1})\equiv\l_k.
\ee
Since the quasi-particle picture is not sensitive to $A_1\cup A_2$ not being a pure state. For our purpose, it's sufficient to compute the charged moments of $\rho_{GGE}$
\be
\begin{split}
&\Tr[\rho_{GGE}^ne^{i\m Q_A}]= Z^{-n}\prod_k\big|\Tr[e^{-(n\l_k-i\m)a_k^{\dg}a_k}]\big|^2\\
&=Z^{-n}\prod_k\big|\sum_{n_k=0}^{\infty}e^{-(n\l_k-i\m)n_k}\big|^2=\prod_k\Big|\frac{(1-e^{-\l_k})^n}{1-e^{-(n\l_k-i\m)}}\Big|^2\\
&=\prod_k|(1+n_k)^n-e^{i\m}n_k^n|^{-2}.
\end{split}
\ee
We have find the density of the logarithmic charged moment in momentum space
\be
\e_{n,\m}(k)=-2\log|(1+n_k)^n-e^{i\m}n_k^n|.
\ee
It's easy to derive that
\be
n_k=\frac{1}{4}\left(\frac{e_k}{e_{0,k}}+\frac{e_{0,k}}{e_k}\right)-\frac12.
\ee
\begin{figure}
        \centering
        \subfloat
        {\includegraphics[width=7cm]{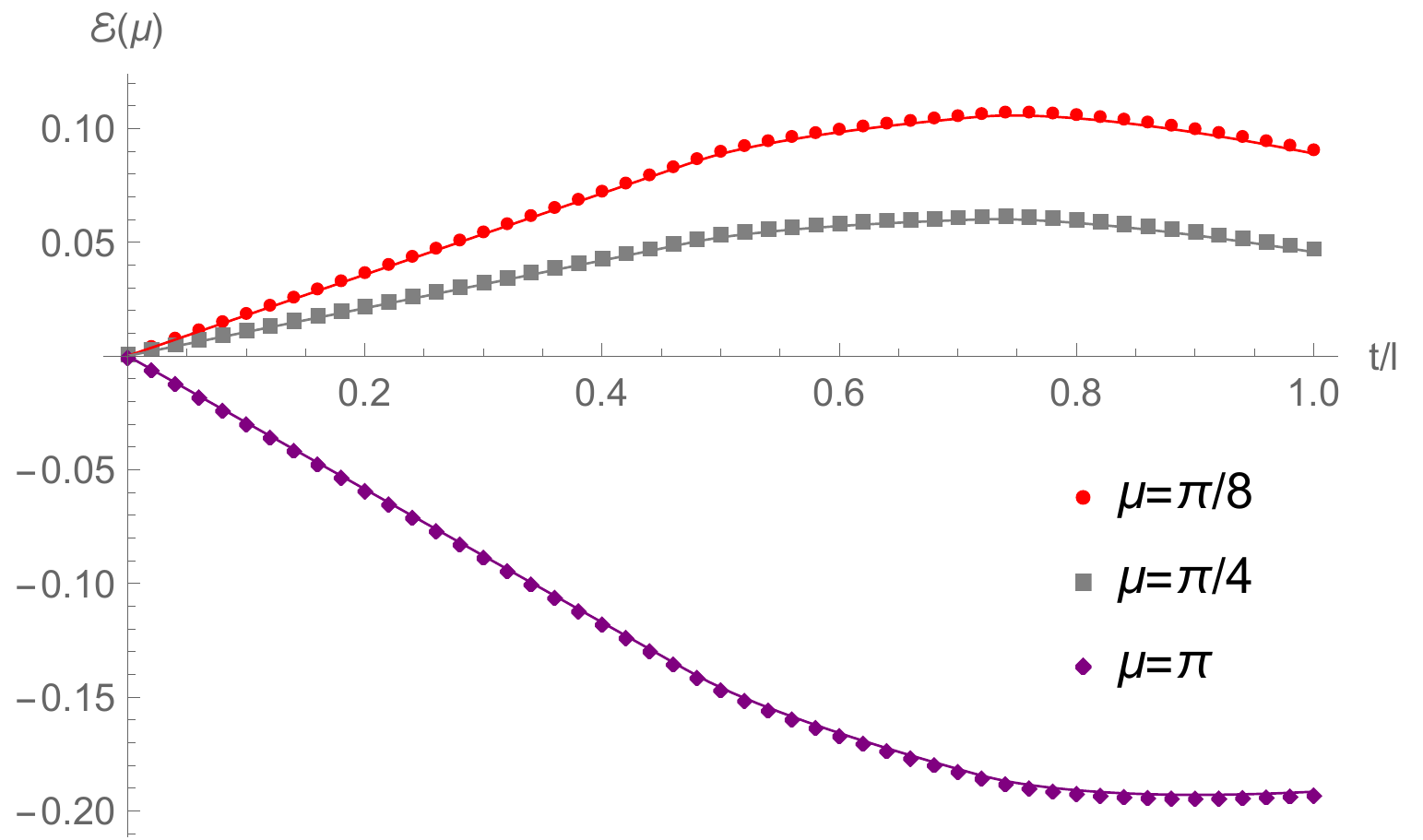}} \quad\quad
        {\includegraphics[width=7cm]{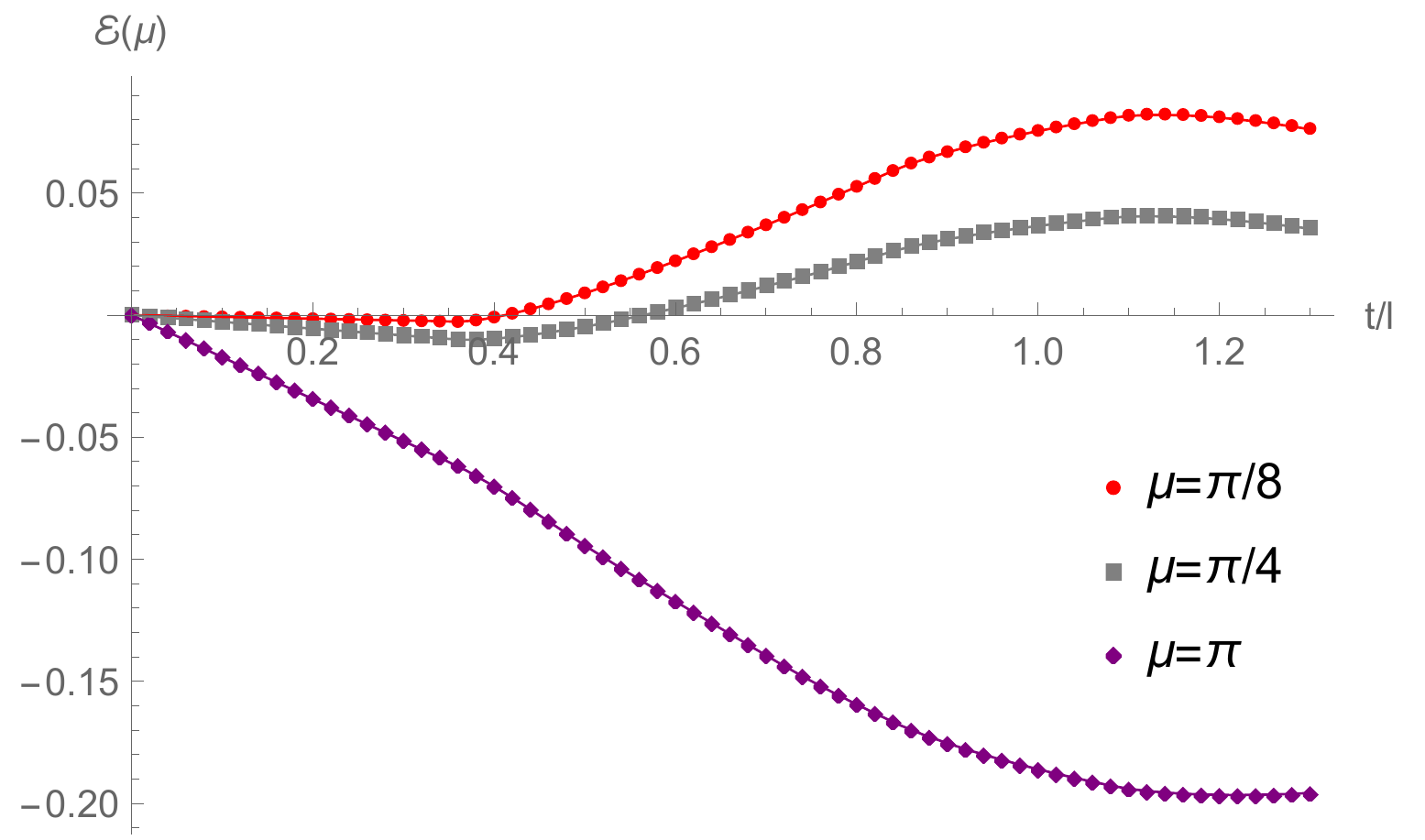}}
        \caption{Charged logarithmic negativity $\mathcal{E}(\m)$ as a function of $t/l$ for different $\m$ after the mass quench from $m_0=1$ to $m=2$ in the complex harmonic chain. The full lines are the quasi-particle predictions (cf. eq.~(\ref{EmiuMa}) and eq.~(\ref{EmiuMd})). Left panel: Adjacent intervals with $L=2500,l_1=200,l_2=300$ and for $\m=\frac{\pi}{8},\frac{\pi}{4},\pi$. Right panel: Disjoint intervals with $L=2500,l_1=200,l_2=300,d=150$ and for $\m=\frac{\pi}{8},\frac{\pi}{4},\pi$. As shown in the figure, the agreement is perfect.}
        \label{fig4}
\end{figure}
Then we have all the ingredients to give conjectures about the dynamics of charged R\'enyi (logarithmic) negativity after a global mass quench in our free lattice model.\\
\\
\pmb{Adjacent interval}
\\
\par For the adjacent intervals, eq.~(\ref{Enmiua}) suggest the following formula of quench dynamics of charged R\'enyi logarithmic negativity
\be\label{EnmiuMa}
\mathcal{E}_n(\m)=\int\frac{dp}{2\pi}\left[\e_{n,\m}^{(2)}(p)/2(\min(2|v_p|t,l_1)+\min(2|v_p|t,l_2))-(\e_{n,\m}^{(2)}(p)/2-\e_{n,\m}(p))\min(2|v_p|t,l)\right].
\ee
The charged logarithmic negativity is obtained by taking the limit $n_e\rightarrow 1$ in $\mathcal{E}_{n_e}(\m)$
\be\label{EmiuMa}
\mathcal{E}(\m)=\int\frac{dp}{2\pi}\left[\e_{1/2,\m}(p)(\min(2|v_p|t,l_1)+\min(2|v_p|t,l_2))-(\e_{1/2,\m}(p)-\e_{1,\m}(p))\min(2|v_p|t,l)\right].
\ee
\\
where we have defined
\be
\e_{n,\m}^{(2)}(p)=
\begin{cases}
\e_{n,2\m}(p),\quad \text{odd}~n\\
2\e_{\frac{n}{2},\m}(p),\quad \text{even}~n
\end{cases}
\ee
\pmb{Disjoint intervals}
\\
\par In this case, according to eq.~(\ref{Enmiud}), we conjecture that the time evolution of charged R\'enyi logarithmic negativity after a global mass quench is given by
\be\label{EnmiuMd}
\begin{split}
&\mathcal{E}_n(\m)=\int\frac{dp}{2\pi}\big[\e_{n,\m}(p)\left(\min(2|v_p|t,l_1)+\min(2|v_p|t,l_2)\right)+(\e_{n,\m}^{(2)}(p)/2-\e_{n,\m}(p))\\
&\times(\max(2|v_p|t,l+d)+\max(2|v_p|t,d)-\max(2|v_p|t,l_1+d)-\max(2|v_p|t,l_2+d))\big].
\end{split}
\ee
The charged logarithmic negativity is obtained by taking the limit $n_e\rightarrow 1$ in $\mathcal{E}_{n_e}(\m)$
\be\label{EmiuMd}
\begin{split}
&\mathcal{E}(\m)=\int\frac{dp}{2\pi}\big[\e_{1,\m}(p)\left(\min(2|v_p|t,l_1)+\min(2|v_p|t,l_2)\right)+(\e_{1/2,\m}(p)-\e_{1,\m}(p))\\
&\times(\max(2|v_p|t,l+d)+\max(2|v_p|t,d)-\max(2|v_p|t,l_1+d)-\max(2|v_p|t,l_2+d))\big].
\end{split}
\ee
The comparisons between numerical computations and the quasi-particle predictions are shown in fig. \ref{fig3} and fig. \ref{fig4}, finding perfect agreement.
\section{Conclusion}\label{section8}
In this paper, we discuss the temporal evolution of charge imbalance resolved negativity after two types of global quenches. Firstly, we have considered the boundary state quench in which the post-quench dynamics is governed by a conformal Hamiltonian. Evaluating the correlators of the fluxed twist fields in the UHP, we obtained the quench dynamics of charged R\'enyi negativity. Then the charge resolved negativity is obtained by Fourier transformation. The total negativity can be reconstructed from these charge resolved ones. We also discussed the mass quench in the underlying lattice model and made conjectures based on the quasi-particle picture.
\par It would be very interesting to investigate the evolution of other charge resolved entanglement measures in the same quenches discussed in this paper, or studying other quench setups such as local joining quench or local operator quenches in field theories and in holographic theories. One may find the breakdown of quasi-particle picture in certain circumstance \cite{Kudler-Flam:2020xqu, Bertini:2022yan}. We will turn to these problems in the future work.
\section*{Acknowledgments}
This work was supported  by the National Natural Science Foundation of China, Grant No.\ 12005081.
\bibliography{2022}

\begin{thebibliography}{10}

\bibitem{Amico:2007ag}
L.~Amico, R.~Fazio, A.~Osterloh, and V.~Vedral, ``{Entanglement in many-body
  systems},'' {\em Rev. Mod. Phys.}, vol.~80, pp.~517--576, 2008.

\bibitem{Calabrese:2009qy}
P.~Calabrese and J.~Cardy, ``{Entanglement entropy and conformal field
  theory},'' {\em J. Phys. A}, vol.~42, p.~504005, 2009.

\bibitem{Eisert:2008ur}
J.~Eisert, M.~Cramer, and M.~B. Plenio, ``{Area laws for the entanglement
  entropy - a review},'' {\em Rev. Mod. Phys.}, vol.~82, pp.~277--306, 2010.

\bibitem{Nishioka:2009un}
T.~Nishioka, S.~Ryu, and T.~Takayanagi, ``{Holographic Entanglement Entropy: An
  Overview},'' {\em J. Phys. A}, vol.~42, p.~504008, 2009.

\bibitem{Ryu:2006bv}
S.~Ryu and T.~Takayanagi, ``{Holographic derivation of entanglement entropy
  from AdS/CFT},'' {\em Phys. Rev. Lett.}, vol.~96, p.~181602, 2006.

\bibitem{Hawking:1974sw}
S.~W. Hawking, ``{Particle Creation by Black Holes},'' {\em Commun. Math.
  Phys.}, vol.~43, pp.~199--220, 1975.
\newblock [Erratum: Commun.Math.Phys. 46, 206 (1976)].

\bibitem{Hawking:1976ra}
S.~W. Hawking, ``{Breakdown of Predictability in Gravitational Collapse},''
  {\em Phys. Rev. D}, vol.~14, pp.~2460--2473, 1976.

\bibitem{Almheiri:2020cfm}
A.~Almheiri, T.~Hartman, J.~Maldacena, E.~Shaghoulian, and A.~Tajdini, ``{The
  entropy of Hawking radiation},'' 6 2020.

\bibitem{Calabrese:2004eu}
P.~Calabrese and J.~L. Cardy, ``{Entanglement entropy and quantum field
  theory},'' {\em J. Stat. Mech.}, vol.~0406, p.~P06002, 2004.

\bibitem{Peres:1996dw}
A.~Peres, ``{Separability criterion for density matrices},'' {\em Phys. Rev.
  Lett.}, vol.~77, pp.~1413--1415, 1996.

\bibitem{vidal2002computable}
G.~Vidal and R.~F. Werner, ``Computable measure of entanglement,'' {\em
  Physical Review A}, vol.~65, no.~3, p.~032314, 2002.

\bibitem{plenio2005logarithmic}
M.~B. Plenio, ``Logarithmic negativity: a full entanglement monotone that is
  not convex,'' {\em Physical review letters}, vol.~95, no.~9, p.~090503, 2005.

\bibitem{calabrese2012entanglement}
P.~Calabrese, J.~Cardy, and E.~Tonni, ``Entanglement negativity in quantum
  field theory,'' {\em Physical review letters}, vol.~109, no.~13, p.~130502,
  2012.

\bibitem{calabrese2013entanglement}
P.~Calabrese, J.~Cardy, and E.~Tonni, ``Entanglement negativity in extended
  systems: a field theoretical approach,'' {\em Journal of Statistical
  Mechanics: Theory and Experiment}, vol.~2013, no.~02, p.~P02008, 2013.

\bibitem{kulaxizi2014conformal}
M.~Kulaxizi, A.~Parnachev, and G.~Policastro, ``Conformal blocks and negativity
  at large central charge,'' {\em Journal of High Energy Physics}, vol.~2014,
  no.~9, pp.~1--25, 2014.

\bibitem{bianchini2016branch}
D.~Bianchini and O.~A. Castro-Alvaredo, ``Branch point twist field correlators
  in the massive free boson theory,'' {\em Nuclear Physics B}, vol.~913,
  pp.~879--911, 2016.

\bibitem{blondeau2016universal}
O.~Blondeau-Fournier, O.~A. Castro-Alvaredo, and B.~Doyon, ``Universal scaling
  of the logarithmic negativity in massive quantum field theory,'' {\em Journal
  of Physics A: Mathematical and Theoretical}, vol.~49, no.~12, p.~125401,
  2016.

\bibitem{Castro-Alvaredo:2019irt}
O.~A. Castro-Alvaredo, C.~De~Fazio, B.~Doyon, and I.~M. Sz\'ecs\'enyi,
  ``{Entanglement content of quantum particle excitations. Part II.
  Disconnected regions and logarithmic negativity},'' {\em JHEP}, vol.~11,
  p.~058, 2019.

\bibitem{Chaturvedi:2016rcn}
P.~Chaturvedi, V.~Malvimat, and G.~Sengupta, ``{Holographic Quantum
  Entanglement Negativity},'' {\em JHEP}, vol.~05, p.~172, 2018.

\bibitem{Malvimat:2017yaj}
V.~Malvimat and G.~Sengupta, ``{Entanglement negativity at large central
  charge},'' {\em Phys. Rev. D}, vol.~103, no.~10, p.~106003, 2021.

\bibitem{chaturvedi2018holographic}
P.~Chaturvedi, V.~Malvimat, and G.~Sengupta, ``Holographic quantum entanglement
  negativity,'' {\em Journal of High Energy Physics}, vol.~2018, no.~5,
  pp.~1--15, 2018.

\bibitem{kudler2019entanglement}
J.~Kudler-Flam and S.~Ryu, ``Entanglement negativity and minimal entanglement
  wedge cross sections in holographic theories,'' {\em Physical Review D},
  vol.~99, no.~10, p.~106014, 2019.

\bibitem{Belin:2013uta}
A.~Belin, L.-Y. Hung, A.~Maloney, S.~Matsuura, R.~C. Myers, and T.~Sierens,
  ``{Holographic Charged Renyi Entropies},'' {\em JHEP}, vol.~12, p.~059, 2013.

\bibitem{Goldstein:2017bua}
M.~Goldstein and E.~Sela, ``{Symmetry-resolved entanglement in many-body
  systems},'' {\em Phys. Rev. Lett.}, vol.~120, no.~20, p.~200602, 2018.

\bibitem{PhysRevLett.121.150501}
H.~Barghathi, C.~M. Herdman, and A.~Del~Maestro, ``R\'enyi generalization of
  the accessible entanglement entropy,'' {\em Phys. Rev. Lett.}, vol.~121,
  p.~150501, Oct 2018.

\bibitem{Murciano:2019wdl}
S.~Murciano, G.~Di~Giulio, and P.~Calabrese, ``{Symmetry resolved entanglement
  in gapped integrable systems: a corner transfer matrix approach},'' {\em
  SciPost Phys.}, vol.~8, p.~046, 2020.

\bibitem{Murciano:2020vgh}
S.~Murciano, G.~Di~Giulio, and P.~Calabrese, ``{Entanglement and symmetry
  resolution in two dimensional free quantum field theories},'' {\em JHEP},
  vol.~08, p.~073, 2020.

\bibitem{Horvath:2020vzs}
D.~X. Horv\'ath and P.~Calabrese, ``{Symmetry resolved entanglement in
  integrable field theories via form factor bootstrap},'' {\em JHEP}, vol.~11,
  p.~131, 2020.

\bibitem{Chen:2021pls}
H.-H. Chen, ``{Symmetry decomposition of relative entropies in conformal field
  theory},'' {\em JHEP}, vol.~07, p.~084, 2021.

\bibitem{Capizzi:2021zga}
L.~Capizzi and P.~Calabrese, ``{Symmetry resolved relative entropies and
  distances in conformal field theory},'' {\em JHEP}, vol.~10, p.~195, 2021.

\bibitem{Capizzi:2021kys}
L.~Capizzi, D.~X. Horv\'ath, P.~Calabrese, and O.~A. Castro-Alvaredo,
  ``{Entanglement of the $3$-State Potts Model via Form Factor Bootstrap: Total
  and Symmetry Resolved Entropies},'' 8 2021.

\bibitem{Zhao:2020qmn}
S.~Zhao, C.~Northe, and R.~Meyer, ``{Symmetry-resolved entanglement in
  AdS$_{3}$/CFT$_{2}$ coupled to U(1) Chern-Simons theory},'' {\em JHEP},
  vol.~07, p.~030, 2021.

\bibitem{Weisenberger:2021eby}
K.~Weisenberger, S.~Zhao, C.~Northe, and R.~Meyer, ``{Symmetry-resolved
  entanglement for excited states and two entangling intervals in
  AdS${}_3$/CFT${}_2$},'' 8 2021.

\bibitem{zhao2022charged}
S.~Zhao, C.~Northe, K.~Weisenberger, and R.~Meyer, ``Charged moments in $ w\_3
  $ higher spin holography,'' {\em arXiv preprint arXiv:2202.11111}, 2022.

\bibitem{Calabrese:2020bys}
P.~Calabrese, J.~Dubail, and S.~Murciano, ``{Symmetry-resolved entanglement
  entropy in Wess-Zumino-Witten models},'' {\em JHEP}, vol.~21, p.~067, 2020.

\bibitem{Capizzi:2022jpx}
L.~Capizzi, O.~A. Castro-Alvaredo, C.~De~Fazio, M.~Mazzoni, and
  L.~Santamar\'\i{}a-Sanz, ``{Symmetry Resolved Entanglement of Excited States
  in Quantum Field Theory I: Free Theories, Twist Fields and Qubits},'' 3 2022.

\bibitem{Ghasemi:2022jxg}
M.~Ghasemi, ``{Universal Thermal Corrections to Symmetry-Resolved Entanglement
  Entropy and Full Counting Statistics},'' 3 2022.

\bibitem{cornfeld2018imbalance}
E.~Cornfeld, M.~Goldstein, and E.~Sela, ``Imbalance entanglement: Symmetry
  decomposition of negativity,'' {\em Physical Review A}, vol.~98, no.~3,
  p.~032302, 2018.

\bibitem{Murciano:2021djk}
S.~Murciano, R.~Bonsignori, and P.~Calabrese, ``{Symmetry decomposition of
  negativity of massless free fermions},'' {\em SciPost Phys.}, vol.~10, no.~5,
  p.~111, 2021.

\bibitem{Chen:2021nma}
H.-H. Chen, ``{Charged R\'enyi negativity of massless free bosons},'' {\em
  JHEP}, vol.~02, p.~117, 2022.

\bibitem{feldman2019dynamics}
N.~Feldman and M.~Goldstein, ``Dynamics of charge-resolved entanglement after a
  local quench,'' {\em Physical Review B}, vol.~100, no.~23, p.~235146, 2019.

\bibitem{Parez:2020vsp}
G.~Parez, R.~Bonsignori, and P.~Calabrese, ``{Quasiparticle dynamics of
  symmetry-resolved entanglement after a quench: Examples of conformal field
  theories and free fermions},'' {\em Phys. Rev. B}, vol.~103, no.~4,
  p.~L041104, 2021.

\bibitem{Parez:2022xur}
G.~Parez, R.~Bonsignori, and P.~Calabrese, ``{Dynamics of
  charge-imbalance-resolved entanglement negativity after a quench in a
  free-fermion model},'' 2 2022.

\bibitem{Scopa:2022gfw}
S.~Scopa and D.~X. Horv\'ath, ``{Exact hydrodynamic description of
  symmetry-resolved R\'enyi entropies after a quantum quench},'' 5 2022.

\bibitem{Calabrese:2005in}
P.~Calabrese and J.~L. Cardy, ``{Evolution of entanglement entropy in
  one-dimensional systems},'' {\em J. Stat. Mech.}, vol.~0504, p.~P04010, 2005.

\bibitem{Hartman:2013qma}
T.~Hartman and J.~Maldacena, ``{Time Evolution of Entanglement Entropy from
  Black Hole Interiors},'' {\em JHEP}, vol.~05, p.~014, 2013.

\bibitem{Calabrese:2016xau}
P.~Calabrese and J.~Cardy, ``{Quantum quenches in 1 + 1 dimensional conformal
  field theories},'' {\em J. Stat. Mech.}, vol.~1606, no.~6, p.~064003, 2016.

\bibitem{Asplund:2015eha}
C.~T. Asplund, A.~Bernamonti, F.~Galli, and T.~Hartman, ``{Entanglement
  Scrambling in 2d Conformal Field Theory},'' {\em JHEP}, vol.~09, p.~110,
  2015.

\bibitem{Nozaki:2013vta}
M.~Nozaki, T.~Numasawa, A.~Prudenziati, and T.~Takayanagi, ``{Dynamics of
  Entanglement Entropy from Einstein Equation},'' {\em Phys. Rev. D}, vol.~88,
  no.~2, p.~026012, 2013.

\bibitem{Coser:2014gsa}
A.~Coser, E.~Tonni, and P.~Calabrese, ``{Entanglement negativity after a global
  quantum quench},'' {\em J. Stat. Mech.}, vol.~1412, no.~12, p.~P12017, 2014.

\bibitem{Wen:2015qwa}
X.~Wen, P.-Y. Chang, and S.~Ryu, ``{Entanglement negativity after a local
  quantum quench in conformal field theories},'' {\em Phys. Rev. B}, vol.~92,
  no.~7, p.~075109, 2015.

\bibitem{Cotler:2016acd}
J.~S. Cotler, M.~P. Hertzberg, M.~Mezei, and M.~T. Mueller, ``{Entanglement
  Growth after a Global Quench in Free Scalar Field Theory},'' {\em JHEP},
  vol.~11, p.~166, 2016.

\bibitem{MohammadiMozaffar:2018vmk}
M.~R. Mohammadi~Mozaffar and A.~Mollabashi, ``{Entanglement Evolution in
  Lifshitz-type Scalar Theories},'' {\em JHEP}, vol.~01, p.~137, 2019.

\bibitem{Zhang:2019kwu}
J.~Zhang and P.~Calabrese, ``{Subsystem distance after a local operator
  quench},'' {\em JHEP}, vol.~02, p.~056, 2020.

\bibitem{Murciano:2021zvs}
S.~Murciano, V.~Alba, and P.~Calabrese, ``{Quench dynamics of R\'enyi
  negativities and the quasiparticle picture},'' 10 2021.

\bibitem{Calabrese:2014yza}
P.~Calabrese, J.~Cardy, and E.~Tonni, ``{Finite temperature entanglement
  negativity in conformal field theory},'' {\em J. Phys. A}, vol.~48, no.~1,
  p.~015006, 2015.

\bibitem{2014Entanglement}
P.~Calabrese, F.~Essler, and A.~M. L?Uchli, ``Entanglement entropies of the
  quarter filled hubbard model,'' {\em Journal of Statistical Mechanics Theory
  and Experiment}, vol.~2014, no.~9, 2014.

\bibitem{Hoogeveen:2014bqa}
M.~Hoogeveen and B.~Doyon, ``{Entanglement negativity and entropy in
  non-equilibrium conformal field theory},'' {\em Nucl. Phys. B}, vol.~898,
  pp.~78--112, 2015.

\bibitem{vidmar2016generalized}
L.~Vidmar and M.~Rigol, ``Generalized gibbs ensemble in integrable lattice
  models,'' {\em Journal of Statistical Mechanics: Theory and Experiment},
  vol.~2016, no.~6, p.~064007, 2016.

\bibitem{Kudler-Flam:2020xqu}
J.~Kudler-Flam, Y.~Kusuki, and S.~Ryu, ``{The quasi-particle picture and its
  breakdown after local quenches: mutual information, negativity, and reflected
  entropy},'' {\em JHEP}, vol.~03, p.~146, 2021.

\bibitem{Bertini:2022yan}
B.~Bertini, K.~Klobas, V.~Alba, G.~Lagnese, and P.~Calabrese, ``{Growth of
  R\'enyi Entropies in Interacting Integrable Models and the Breakdown of the
  Quasiparticle Picture},'' 3 2022.

\end{thebibliography}
\bibliographystyle{ieeetr}
\end{document}